\documentstyle[12pt]{article}
\begin{document}

\begin{titlepage}

\begin{center}
{\bf HIGH-ORDER HARMONIC GENERATION IN LASER-IRRADIATED
HOMONUCLEAR DIATOMICS: THE VELOCITY GAUGE VERSION OF MOLECULAR
STRONG-FIELD APPROXIMATION}
\\[30pt]

{\bf Vladimir I. Usachenko}{\it \footnote[1]{{\it Corresponding
author, permanent mailing address: Institute of Applied Laser
Physics UzAS, Katartal str., 28, Tashkent, 700135, Uzbekistan.
E-mail: vusach@yahoo.com.}}}$^{,2,3}${\bf , Pavel E. Pyak$^{2}$
and Shih-I Chu}$^{4}$
\\[10pt]

$^{1}${\it Institute of Applied Laser Physics UzAS, Katartal
str. 28, Tashkent, 100135, Uzbekistan,}\\[3pt]
$^{2}${\it Physics Department, National University of Uzbekistan,
Tashkent, 100174, Uzbekistan}\\[3pt]
$^{3}${\it Max-Born-Institute for Nonlinear Optics and Short-Pulse
Laser Spectroscopy,}\\
{\it Berlin, 12489, Germany}\\[3pt]
$^{4}${\it Department of Chemistry, University of Kansas,
Lawrence, KS 66045-7582, USA}\\[30pt]
\end{center}

\noindent {\rm The generation of high harmonics in
laser-irradiated light homonuclear diatomics (}$H_2^{+}${\rm ,
}$N_2$ {\rm and }$O_2${\rm ) compared to that in atomic
counterparts (of nearly identical binding energy) is studied
within the {\it velocity gauge} version of conventional {\it
strong-field approximation}. The applied strong-field approach
(alternatively developed earlier to incorporate rescattering
effects beyond the conventional saddle-point approximation) is
currently extended to molecular case by means of supplement the
standard {\it linear combination of atomic orbitals} and {\it
molecular orbitals} method. The associated model proved to
adequately reproduce a general shape and detailed structure of
molecular harmonic spectra, which demonstrate a number of
remarkable distinctive differences from respective atomic spectra
calculated under the same laser pulses. The revealed differences
are found to be strongly dependent on internuclear separation and
also very sensitive to the orbital and bonding symmetry of
contributing molecular valence shell. In particular, the model
correctly predicts the behavior of high-frequency plateau (both
for its extent and even details of structure) in molecular
harmonic spectra at small (nearly equilibrium) and large
internuclear separations. In addition, for some group of
harmonics, the harmonic emission rates were ascertained to
dominate by contribution from inner molecular shells of higher
binding energy and different orbital symmetry compared to the
outermost molecular orbital normally predominantly contributing.}
\end{titlepage}

\begin{center}
{\bf HIGH-ORDER HARMONIC GENERATION IN LASER-IRRADIATED
HOMONUCLEAR DIATOMICS: THE VELOCITY GAUGE VERSION OF MOLECULAR
STRONG-FIELD APPROXIMATION}
\\[30pt]

{\bf Vladimir I. Usachenko}{\it \footnote[1]{{\it Corresponding
author, permanent mailing address: Institute of Applied Laser
Physics UzAS, Katartal str., 28, Tashkent, 700135, Uzbekistan.
E-mail: vusach@yahoo.com.}}}$^{,2,3}${\bf , Pavel E. Pyak$^{2}$
and Shih-I Chu}$^{4}$
\\[10pt]

$^{1}${\it Institute of Applied Laser Physics UzAS, Katartal
str. 28, Tashkent, 100135, Uzbekistan,}\\[3pt]
$^{2}${\it Physics Department, National University of Uzbekistan,
Tashkent, 100174, Uzbekistan}\\[3pt]
$^{3}${\it Max-Born-Institute for Nonlinear Optics and Short-Pulse
Laser Spectroscopy,}\\
{\it Berlin, 12489, Germany}\\[3pt]
$^{4}${\it Department of Chemistry, University of Kansas,
Lawrence, KS 66045-7582, USA}\\[30pt]
\end{center}

\begin{enumerate}
\item  {\rm \noindent INTRODUCTION AND BACKGROUND MOTIVATION.}

{\rm \qquad Apart from pure fundamental interest, the strong-field
phenomenon of high-order harmonic generation (HHG) in
laser-irradiated species is currently believed to be a highly
promising for creation compact (table-top) sources of ultra-short
(sub-femtosecond) pulses of powerful coherent XUV radiation [1,
2]. Being accessible, such a high-frequency radiation might be
used for X-ray holography or monitoring of biological processes
{\it in vivo} [2] as well as tomographic imaging of molecular
orbitals [3] with an unprecedented brightness and spatial and
temporal resolution. For the reasons above, the HHG process became
a subject of equally great experimental and theoretical interest
in the last decade (see also e.g., [4-6], for review and
references cited therein). Compared to atoms, the laser-irradiated
molecules as sources of high harmonics [7-14] offer promising
prospects as to raising the up to now disappointingly low
conversion efficiency for harmonic generation [7-12] and/or
harmonic cutoff frequency [10, 13] due to an advanced possibility
of controlling and manipulating the additional external degrees of
freedom in molecules. For example, the internuclear distance
}$R_0$ {\rm of simplest diatomic molecules, which determines
differently delocalized initial electronic states, can be used to
enhance the harmonic conversion efficiency exceeding the atomic
one due to transient enhancement of harmonic emission in expanding
molecules [12]. Particularly, the calculated harmonic conversion
efficiency was found to be noticeably enhanced (by several orders
of magnitude) with increasing of }$R_0$ {\rm over equilibrium
internuclear separation [10, 12, 14]. Diatomic molecules may also
produce high harmonics of much higher cutoff frequency due to a
much longer high-frequency plateau extended up to }$8U_p$ {\rm of
maximum length [10] versus the respective maximum value }$3.17U_p$
{\rm ascertained for atomic harmonic spectra (here }$U_p${\rm is
the ponderomotive energy of field-induced oscillating motion of an
electron in the continuum; the {\it atomic system of units} is
used unless stated otherwise).}

{\rm \qquad Because of the electron wavelength is comparable to
the size of the molecular wavefunction, the intensity of emission
at a particular X-ray photon energy is also expected to be
sensitive to the wavelength of the electron wave and the shape of
the molecular wavefunction and/or orientation of internuclear
molecular axis with respect to incident laser field polarization
[8-11]. In addition, simultaneous observation of both ion yields
and harmonic signals [15] under the same conditions was shown to
serve as an effective tool to probe molecular structure in an
instant [16] and also make it possible to use quantum phenomena in
HHG associated with molecular symmetries to disentangle the
contributions from the ionization and recombination processes
[17]. In this context, homonuclear (or homopolar) diatomics seem
to be also of special interest due to identical atomic nuclei give
rise to a possibility of various laser field-induced and
orientation-dependent {\it intramolecular interference} effects
[15] that may additionally affect the molecular HHG process. The
clear signs of intramolecular interference-related effects due to
two-centered nature of diatomic molecule were numerically revealed
[11] in high-harmonic spectra of model (}$1D${\rm \ or }$2D${\rm )
}$H_2^{+}${\rm \ and }$H_2${\rm \ systems as a broad and
pronounced local minimum of the position strongly dependent on the
internuclear separation and spatial orientation of diatomic
molecule. The origin of the minimum was attributed to the {\it
destructive} interference of photoelectron emission from separate
atomic centers [11], so that a transparent interpretation can be
given in terms of interference between two radiating point sources
located at the positions of the nuclei. The latter pivotal idea
(obviously proposed first in Ref. [18]) about intramolecular
interference of atomic photoionization amplitudes was successfully
exploited in [19] for description of molecular ionization as
compared to ionization of respective atomic counterparts having
nearly identical binding energies. Based on the {\it velocity
gauge} formulation of conventional {\it strong-field
approximation} (SFA) (see, e.g. [20] and associated {\it
Keldysh-Faisal-Reiss} theories [21]) with the atomic ionization
rates modified by the interference from the atomic centers, the
molecular ionization was predicted to be enhanced (or, at least,
not suppressed) in }$ N_2$ ({\rm versus }$Ar${\rm ) and/or
suppressed in }$O_2$ {\rm (versus }$Xe$ {\rm ). The mentioned
suppression in }$O_2${\rm \ was interpreted in terms of the
strong-field MO-SFA model [19] as intimately related to
intramolecular interference, which is to be always destructive for
low-energy photoelectrons emitted from the }$1\pi _g$ {\it
highest-occupied molecular orbital }{\rm (HOMO) of {\it
antibonding} symmetry and predominantly contributing to respective
ionization rate. Accordingly, the background physical mechanism
underlying an enhanced ionization in }$N_2$ {\rm versus }$ Ar$
{\rm was identified as constructive intramolecular interference
for low-energy photoelectrons emitted from the }$3\sigma _g$ {\rm
HOMO in }$N_2$ {\rm of {\it bonding} symmetry. }

{\rm \qquad Meantime, because of the high suppression in
strong-field ionization of }$O_2${\rm \ versus }$Xe${\rm , the
laser-irradiated diatomic } $O_2$ {\rm is expected to produce high
harmonics of noticeably higher {\it cutoff} frequency as compared
to }$Xe$ {\rm atom [13]. Namely, due to a suppressed molecular
ionization in }$H_2$ {\rm and }$O_2$ {\rm (versus their atomic
counterparts) and correspondingly larger value of the saturation
intensity, the latter diatomics} {\rm survive at a considerably
higher laser intensity (at which their atomic counterparts are to
be completely ionized) and, thus, they are still able to produce
harmonics of a considerably higher {\it cutoff} frequency
(proportional to laser intensity). The related extension of
high-frequency plateau in harmonic spectra of }$O_2$ {\rm relative
to }$Xe${\rm , was recently observed in experiment and numerically
simulated [13] based on the Lewenstein model [22] supplemented by
the molecular tunneling ionization (or the so-called MO-ADK)
theory [23] to account for the differences in ionization of
molecules compared to atoms in terms of {\it
Ammosov-Delone-Krainov} (ADK) model of tunneling ionization [24].
The MO-ADK model partly succeeded in reproducing of suppressed
ionization (e.g., in }$O_2$ {\rm versus }$Xe$ {\rm compatible with
experiment [25]), although, being applied to molecular HHG, the
model seems to encounter difficulties in providing a transparent
interpretation and satisfactory agreement with experiment in
prediction of the {\it cutoff} frequency observed in molecular
high-harmonic spectra [13] (see also Sec.4 below, for details).
Being also applied to ionization of other diatomic species (e.g.,
in }$H_2${\rm \ versus companion }$Ar$ {\rm atom), the ADK-based
approach resulted in a discrepancy with experiment [26], at least
in the strong-field domain (i.e., only where the tunneling theory
is generally valid). Since the wavefunction of initial molecular
state corresponding to HOMO is always approximated by only a {\it
single one-centered} atomic orbital within the MO-ADK model, such
theory is not suitable either for adequate description of
intramolecular interference-related phenomena arising from the
{\it two-centered} nature of molecular valence shell in
diatomics.}

\qquad {\rm Thus, there are numerous interrelated and remarkable
aspects of molecular HHG and above-threshold ionization (ATI) as a
prerequisite to the HHG, which appear to be also strongly
interdependent and interrelated in laser-irradiated molecules too.
However, with the possible few exceptions, the fundamental physics
associated with how atoms and molecules respond to intense laser
radiation (i.e. optical fields approaching the atomic unit) cannot
be treated exactly due to the comparable strengths of the
electron-nucleus, electron-electron and electron-field
interaction. Although {\it ab initio} quantum calculations for
atoms are readily available, at least within the{\it \
single-active electron} (SAE) approximation, this is not the case
for molecules as fully three-dimensional }($3D${\rm ) calculations
on even relatively simple diatomics, such as }$H_2^{+}${\rm , are
extremely difficult [14]. Therefore, the HHG process in diatomic
molecules is primarily treated nowadays using approximate
SAE-based general strong-field approaches, such as conventional
}{\it strong-field approximation} {\rm (SFA) formulated in terms
of ''rescattering'' (or recollision) picture and model zero-range
binding potential of atomic centers [10] and/or various pure
numerical procedures (see e.g. [27]) and simulations (e.g., [8,
11, 12]). The harmonic emission rates for molecules can be also
calculated, in principle, based on different ab initio numerical
procedures and methods, such as direct numerical solution of
time-dependent Schr\"{o}dinger equation (TDSE method, e.g., [11])
and time-dependent density-functional theory (TD-DFT) including
many-electron effects [28]; however, the related results seem to
be very computationally demanding and hardly available for
reliable and transparent interpretation.}

\qquad {\rm All the aspects above gave us a strong motivation to
consider the harmonic generation in laser-irradiated diatomics and
verify the mentioned aspects within the currently proposed
alternative model of molecular HHG. The model is based on fully
quantum-mechanical SAE-based strong-field approach developed
earlier [29, 30] and proved to adequately incorporate rescattering
effects in atomic HHG and high-energy ATI processes. In the
present paper the proposed strong-field approach is extended to
molecular case of diatomic specie (i.e. substantially
many-electron system) by means of supplement the standard {\it
linear combination of atomic orbitals} (LCAO) and {\it molecular
orbitals} (MO) method invoked for approximate reproducing the
two-centered one-electron wavefunction of initial molecular
discrete state. The LCAO-MO method proved to be surprisingly well
working for adequate description and transparent interpretation
the phenomena of enhanced (and/or suppressed) ionization of
laser-irradiated diatomics in terms of previously developed
strong-field SFA-LCAO model [31]. This model is to be also used
further in Sec.3 for numerical calculation and evaluation of laser
intensity for ionization saturation under conditions of molecular
HHG experiment [13]. The related results for calculated harmonic
emission rates (or high-harmonic conversion efficiency) and
position of high-frequency plateau cutoff proved to be sensitive
to the molecular orbital symmetry and internuclear separation. At
last, our present consideration is not restricted to only one, a
single (the outermost) molecular valence shell, so that other
(inner) contributing molecular orbitals of higher binding energies
and different orbital symmetry are additionally and equally well
incorporated, wherever appropriate. The proposed molecular HHG
model is shown to adequately reproduce the distinctive features of
molecular harmonic spectra compared to respective atomic spectra
as well as noticeable differences found in structure of molecular
harmonic spectra corresponding to HOMO of bonding symmetry (e.g.,}
$3\sigma _g$\ {\rm in} $N_2${\rm ) relative to antibonding
symmetry (e.g., } $1\pi _g$ {\rm in }$O_2$).{\rm \ }\\

\item  {\rm THE APPLIED STRONG-FIELD HHG MODEL: BASIC ASSUMPTIONS
AND ANALYTICAL RELATIONS.}

{\rm \qquad The harmonic emission process under discussion is to
be further investigated within framework of the strong-field fully
quantum-mechanical approach developed in [29, 30] and applied
previously to treatment of atomic HHG process. Let's outline first
in brief the main features of related strong-field model, which is
currently to be extended to molecular case. Within the {\it dipole
approximation} (neglecting any photon momenta) the Hamiltonian of
EM interaction in the {\it velocity gauge} (VG) form,  reads as}
\begin{equation}
\widehat{W}\left( {\bf r},t\right) =\frac 1{c_0}{\bf A}\left(
t\right) \cdot \widehat{{\bf p}}+\frac 1{2c_0^2}{\bf A}^2\left(
t\right)   \label{1}
\end{equation}
{\rm where }$\widehat{{\bf p}}{\bf =}-i{\bf \nabla }$ {\rm is the
operator of electron canonical momentum, }$c_0\approx 137${\rm \
is the light velocity in vacuum. This means that the vector
potential }${\bf A}(t)${\rm \ of incident laser field and
associated field strength }${\bf E}(t)${\rm \ are supposed to be
independent on coordinate }${\bf r}${\rm , but both are functions
of time }$t${\rm \ only. Our present consideration will be
restricted entirely to the case of only harmonic emission
processes in which the atomic or molecular specie occupies the
same state before and after the passage of the laser pulse. Thus,
the bound discrete state is supposed to be unperturbed by incident
laser field and the respective undistorted (laser-free) stationary
wavefunction }$\Phi _0\left( {\bf r},t\right) =\Phi _0\left( {\bf
r}\right) \exp \left( -i\varepsilon _0t\right) $ {\rm already
known in advance (or, at least, found to any arbitrary prescribed
accuracy).}

\qquad {\rm Then, for particular case of linearly polarized laser
field of }the frequency $\omega $, electric field strength $E$ and
unit polarization vector ${\bf e}${\rm : }
\begin{equation}
{\bf A}_L\left( t\right) =\left( c_0/\omega \right) {\bf e}E\cos
\left( \omega t\right) {\rm \ }  \label{2}
\end{equation}
{\rm \ the total SFA-amplitude of HHG can be expressed through the
sum of partial amplitudes }$f_N^{\left( HHG\right) }\left( \Omega
_{{\bf k}^{\prime }}\right) $ {\rm of emission of high-harmonic of
}$N${\rm -th order, frequency $\Omega _{{\bf k}^{\prime }}$ and
polarization ${\bf e}_{{\bf k}^{\prime },\lambda ^{\prime }}$
($\lambda = 1, 2$):}
\begin{equation}
f_{i\rightarrow f}^{\left( HHG\right) }\left( \Omega _{{\bf k}^
{\prime }}\right) =\sum_Nf_N^{\left( HHG\right) }\left( \Omega
_{{\bf k}^{\prime }}\right) \delta \left( \Omega _{{\bf k}^{\prime
}}-N\omega \right) = \label{3}
\end{equation}
\[
=-\sqrt{\frac{2\pi }{V\Omega _{{\bf k}^{\prime }}}}\int_{-\infty
}^\infty Q_\Omega \left( t\right) \exp \left( i\Omega _{{\bf
k}^{\prime }}t\right) dt
\]
where
\begin{equation}
Q_\Omega \left( t\right) =i\Omega _{{\bf k}^{\prime }}D_\Omega
\left( t\right) =i\Omega _{{\bf k}^{\prime
}}\sum_N\widetilde{D}_\Omega ^{\left( N\right) }\left( \omega
\right) \exp \left( -iN\omega t\right)   \label{4}
\end{equation}
with the time-dependent dipole moment $D_\Omega \left( t\right)$
induced in a laser-irradiated system by incident laser field,
whereas $V$\ is the spatial normalization volume. The $N$-th
Fourier component $\widetilde{D} _\Omega ^{\left( N\right) }\left(
\omega \right) $ of the field-induced dipole moment can be derived
in the following analytical form [29]:
\begin{equation}
\widetilde{D}_\Omega ^{\left( N\right) }\left( \omega \right)
=\sum_{m\geq N_0}R_{N-m,m}\left( \Omega _{{\bf k}^{\prime
}},q_m,\eta \right) = -i \pi \Omega^{-1}_{{\bf k}^{\prime
}}\sum_{m\geq N_0}\left( U_p-m\omega \right) \times   \label{5}
\end{equation}
\[
\times q_m\int \left( {\bf e}_{{\bf k}^{\prime },\lambda ^{\prime
}}^{*}\cdot {\bf q}_m\right) B_{N-m}\left( \zeta ({\bf q}_m);\frac
\eta 2\right) B_{-m}\left( \zeta ({\bf q}_m);\frac \eta 2\right)
\left| F_0\left( {\bf q}_m\right) \right| ^2dO_{{\bf q}_m}
\]
Here
\begin{equation}
F_0\left( {\bf q}\right) \equiv \left( 2\pi \right) ^{-3/2}\int
d{\bf r\cdot }\exp \left( -i{\bf q\cdot r}\right) \Phi _0({\bf r})
\label{6}
\end{equation}
is the Fourier transform of stationary wavefunction $\Phi _0({\bf
r})$ of laser-exposed specie corresponding to initial bound
discrete state unperturbed by incident laser field, while
$B_s\left( x;y\right) $ is {\rm the }$s$-th\ order {\rm
generalized Bessel function [20] of two real arguments as
dimensionless parameters:}
\begin{equation}
\zeta \left( {\bf p}\right) =\left( {\bf E\cdot p}\right)
/\omega^2,\ \ \ \eta =U_p/\omega =E^2/\left( 4\omega ^3\right)
\label{7}
\end{equation}
{\rm with ponderomotive energy }$U_p${\rm \ of field-induced
oscillating motion of electron driven in continuum.} The variable
$q_m=\sqrt{2\left( m\omega -I_p-U_p\right) }$ in Eq.(5) denotes
discrete values of photoelectron canonical momentum in
intermediate continuum states of energies
\begin{equation}
\varepsilon _{{\bf q}}^{(m)}=m\omega -I_p-U_p  \label{8}
\end{equation}
{\rm corresponding to $m$ number of absorbed laser photons
beginning from the minimum one $N_0\equiv N_0\left( \eta \right)
=\left[ \left( I_p+U_p\right) /\omega \right] +1$\ required for
ionization (here $\left[ x\right] $\ denotes an integer part of
variable $x$).}

\qquad {\rm Due to the ''{\it pole approximation}'' applied, the
summation in (5) over positive integer $m\geq N_0$ takes into
account the resonance (or ''{\it essential}'') intermediate
continuum states corresponding to open ATI channels (to which the
process of direct ATI is also possible), which are only supposed
to give the main (predominant) contribution to HHG process under
consideration (see also [29] for details). It is also worth noting
that the time-independent matrix function $R_{n,m}\left( \Omega
_{{\bf k}^{\prime }},q_m,\eta \right) $ allows for representation
in the factorized form:}
\begin{equation}
R_{n,m}\left( \Omega _{{\bf k}^{\prime }},q_m,\eta \right) =\int
f_n^{(SR)}\left( \Omega _{{\bf k}^{\prime }},{\bf q}_m,\eta
\right) f_m^{(MPI)}\left( {\bf q}_m,\eta \right) dO_{{\bf q}_m}
\label{9}
\end{equation}
that makes the final results transparent for interpretation.
Namely, the integrand in (9) is a product of time-independent
amplitudes of two strong-field multiphoton processes - the direct
ATI process accompanied by absorption of $m$ incident laser
photons and emission of a photoelectron to intermediate continuum
states of canonical momentum ${\bf q}_m$
\begin{equation}
f_m^{(MPI)}\left( {\bf q}_m,\eta \right) \sim 2\pi \left(
U_p-m\omega \right) B_{-m}\left( \zeta ({\bf q}_m);\frac \eta
2\right) F_0\left( {\bf q} _m\right)   \label{10}
\end{equation}
followed by subsequent process of spontaneous photorecombination
and emission of a high-harmonic photon of frequency $\Omega _{{\bf
k}^{\prime }}$ and polarization ${\bf e}_{{\bf k}^{\prime
},\lambda ^{\prime }}$ ($\lambda = 1, 2$):
\begin{equation}
f_n^{(SR)}\left( \Omega _{{\bf k}^{\prime }},{\bf q}_m,\eta
\right) \sim iq_m \sqrt{2\pi /\Omega^3_{{\bf k}^{\prime }}}\left(
{\bf e}_{{\bf k}^{\prime },\lambda ^{\prime }}^{*}\cdot {\bf
q}_m\right) B_{n}\left( \zeta ({\bf q} _m);\frac \eta 2\right)
F_0^{*}\left( {\bf q}_m\right)   \label{11}
\end{equation}
{\rm Finally, the high-harmonic spectrum within the proposed model
is represented by a sequence of equidistant (separated by a laser
fundamental frequency $\omega $) discrete peaks of heights defined
by the partial HHG amplitudes $f_N^{\left( HHG\right) }(\Omega
_{{\bf k}^{\prime }})$ (3) and respective differential rates
$w_N^{\left( HHG\right) }({\bf k}^{\prime })$:}
\begin{equation}
w_N^{\left( HHG\right) }({\bf k}^{\prime })= \frac{ \Omega^3
_N}{\left( 2\pi c_0\right) ^3}\left| \widetilde{D}^{\left(
N\right)}_{\Omega_{N}}\left( \omega \right) \right| ^2  \label{12}
\end{equation}
{\rm of emission of high-harmonic photon of discrete frequency
$\Omega _N=N\omega $ to a fixed solid angle element }$dO_{{\bf
k}^{\prime }}${\rm \ . }

\qquad {\rm The total ionization rates $\Gamma _{ion}\left( \eta
\right) $ can be derived just from the amplitude (10) of direct
ATI process by means of integration of its squared modulus over
the angles of photoelectron emission as well as photoelectron
energy. The integration over the final energy of emitted
photoelectron is just reduced to the summation over the total
number $N$ of laser photons absorbed (see, also [30, 31], for
details): }
\begin{equation}
\Gamma _{ion}\left( \eta \right) =\sum_{N\geq N_0\left( \eta
\right) }\frac{p_N}{\left( 2\pi \right) }\left( U_p-N\omega
\right) ^2\int dO_{{\bf p} _N}B_{-N}^2\left( \zeta \left( {\bf
p}_N\right) ;\frac \eta 2\right) \left| F_0\left( {\bf p}_N\right)
\right| ^2  \label{13}
\end{equation}
{\rm In a pulsed laser field of arbitrary duration $\tau $, the
total yield of photoelectrons (i.e., the produced photoelectron
and/or ion signal that is measured in experiment) depends on the
peak field strength $E_0$ and laser's spatial and temporal
profiles. For spatially homogeneous incident laser field, the
amplitude $E$ of laser field is only dependent on the peak field
strength $E_0$ and temporal profile function $g\left( t\right) $
according to}:
\begin{equation}
{\bf E}\left( t\right) ={\bf e}E\left( t\right) \sin \left( \omega
t\right) = {\bf e}E_0g\left( t\right) \sin \left( \omega t\right)
\label{14}
\end{equation}
{\rm So, for example, for {\it Gaussian} shape of laser pulse
intensity: }
\begin{equation}
g\left( t\right) =\exp \left[ -2\left( \frac{t-t_0}\tau \right)
^2\ln 2\right]  \label{15}
\end{equation}
{\rm with $t_0=\tau /2$ as the time moment at which the laser
intensity reaches a maximum during the laser pulse duration. Under
conventional supposition that the ionization process is generally
a considerably faster than the time period of essential change of
temporal profile function $ g\left( t\right) $, the ionization
rate (13) derived assuming time-independent laser field intensity
can be approximately considered as still valid, but corresponding
to the instant ionization rate. This means that the ionization
rate (13) is just supposed to be adiabatically changing with
variation of laser intensity through the laser pulse action and
the total ionization probability for the overall time of laser
pulse duration $\tau $ can be found according to the quasistatic
formula:}
\begin{equation}
P_{ion}(\tau )=1-\exp \left( -\int_0^\tau \Gamma _{ion}\left(\eta
\left( t\right) \right) dt\right)   \label{16}
\end{equation}
{\rm where, particularly}
\begin{equation}
\eta \left( t\right) =E^2\left( t\right) /\left( 4\omega ^3\right)
=g^2\left( t\right) E_0^2/\left( 4\omega ^3\right)   \label{17}
\end{equation}
{\rm The Eq.(16) will be used further for defining the saturation
ionization intensity $I_{sat}$ as the laser intensity which leads
to $99\%$ of the laser-irradiated species being ionized by the
ending time of the
laser pulse, so that $P_{ion}(\tau )\approx 0.99$.}\\

\item  SAE-EXTENSION TO MOLECULAR HHG IN DIATOMIC MOLECULES:
THE MOLECULAR ORBITALS IN COORDINATE AND MOMENTUM SPACE.

\qquad {\rm So far, up to this point, we didn't specify the nature
of laser-irradiated system, so that all the relations above are
supposed to be equally well applicable to arbitrary atomic and/or
molecular species provided the SAE consideration is valid to
derive the Fourier transform of a single-electron wavefunction of
initial bound discrete state in a closed analytical form. Thus,
according to the HHG model currently extended to molecular case of
homonuclear diatomics, the specified properties of
laser-irradiated system are incorporated solely through the
ionization potential }$I_p^{(n)}$ {\rm and Fourier transform (6)
being the wavefunction }$ F_n\left( {\bf p};{\bf R}_0\right) $
{\rm of initial $n$-th\ molecular discrete state in momentum
space. The corresponding single-electron wavefunctions, or {\it
molecular orbitals}, each of a fixed discrete binding energy
}$\varepsilon _0^{\left( n\right) }=-I_p^{(n)}$ {\rm and
respective number }$N_e^{(n)}$ {\rm of identical electrons, are
the mathematical constructs used to describe the multi-electron
wavefunction in molecules (similar to {\it atomic orbitals }in
atom). According to the standard linear combination of atomic
orbitals (LCAO) and molecular orbitals (MO) method, the
wavefunction of each (}$n${\rm -th) molecular valence shell can be
approximately considered as {\it two-centered }MO:}
\[
\Phi _n\left( {\bf r}_1;{\bf r}_2\right) =\Phi _n\left( {\bf
r}-{\bf R}_0/2 {\bf ;r}+{\bf R}_0/2\right) =
\]
\begin{equation}
=\sum_j \sqrt \frac{N_e^{(n)}} {2 \left( 1\pm S_j^{(n)}\left(
R_0\right) \right) } \left[ \phi _j^{(n)}\left( {\bf r}-{\bf
R}_0/2\right) \pm \phi _j^{(n)}\left( {\bf r}+{\bf R}_0/2\right)
\right] \label{18}
\end{equation}
{\rm being a linear superposition of predominantly contributing
(}$j${\rm -th) {\it one-electron }AOs }$\phi _j^{(n)}\left( {\bf
r}\right) ${\rm \ centered on each of the atomic cores and thus
separated by internuclear distance }$R_0${\rm . Here }
\begin{equation}
S_j^{(n)}\left( R_0\right) =\int d{\bf r}\phi _j^{(n)}\left( {\bf
r}+{\bf R} _0/2\right) \phi _j^{(n)}\left( {\bf r}-{\bf
R}_0/2\right)   \label{19}
\end{equation}
{\rm is the respective atomic orbital overlap integral.}

\qquad {\rm Because the highest-lying orbitals are responsible for
chemical properties, they are of particular interest. For
}$N_2${\rm \ molecule and the }$3\sigma _g$ {\rm HOMO of gerade
and bonding symmetry commonly considered, the approximate
two-centered single-electron molecular wavefunction can be
composed, for example, as a superposition of scaled hydrogen-like
}$2p_z${\rm \ }or $1s${\rm \ atomic orbitals [31]:}
\begin{equation}
\Phi _{\left( 2p\right) 3\sigma _g}\left( {\bf r};{\bf R}_0\right)
=\sqrt \frac{N_e^{(3\sigma _g)}} {2 \left( 1-S_{2p_z}^{(3\sigma
_g)}\left( R_0\right) \right) } \left[ \phi _{2p_z}^{(3\sigma
_g)}\left( {\bf r}-{\bf R}_0/2\right) -\phi _{2p_z}^{(3\sigma
_g)}\left( {\bf r}+{\bf R}_0/2\right) \right] \label{20}
\end{equation}
\begin{equation}
\Phi _{\left( 1s\right) 3\sigma _g}\left( {\bf r};{\bf R}_0\right)
=\sqrt \frac{N_e^{(3\sigma _g)}} {2 \left( 1+S_{1s}^{(3\sigma
_g)}\left( R_0\right) \right) }\left[ \phi _{1s}^{(3\sigma
_g)}\left( {\bf r}-{\bf R}_0/2\right) +\phi _{1s}^{(3\sigma
_g)}\left( {\bf r}+{\bf R}_0/2\right) \right]   \label{21}
\end{equation}
{\rm Accordingly, for ungerade }$1\pi _u$ {\rm inner valence shell
of bonding symmetry or gerade }$1\pi _g$ {\rm HOMO of antibonding
symmetry commonly considered in }$O_2${\rm \ molecule,\ the
respective approximate two-centered single-electron molecular
wavefunction can be composed, for example, as a superposition of
scaled hydrogen-like\ }$2p_x${\rm \ (or }$2p_y $){\rm \ atomic
orbitals [31]:}
\begin{equation}
\Phi _{\left( 2p\right) 1\pi _u}\left( {\bf r};{\bf R}_0\right)
=\sqrt \frac{N_e^{(1\pi _u)}} {2 \left( 1+S_{2p_x}^{(1\pi
_u)}\left( R_0\right) \right) } \left[ \phi _{2p_x}^{(1\pi
_u)}\left( {\bf r}-{\bf R}_0/2\right) +\phi _{2p_x}^{(1\pi
_u)}\left( {\bf r}+{\bf R}_0/2\right) \right] \label{22}
\end{equation}
\begin{equation}
\Phi _{\left( 2p\right) 1\pi _g}\left( {\bf r};{\bf R}_0\right) =
\sqrt \frac{N_e^{(1\pi _g)}} {2\left( 1-S_{2p_x}^{(1\pi _g)}\left(
R_0\right) \right) } \left[ \phi _{2p_x}^{(1\pi _g)}\left( {\bf
r}-{\bf R}_0/2\right) -\phi _{2p_x}^{(1\pi _g)}\left( {\bf r}+{\bf
R}_0/2\right) \right] \label{23}
\end{equation}
{\rm Note first that the {\it ''}}$-${\rm {\it '' }combination in
the right-hand side of Eq.(18) doesn't not necessarily correspond
to an {\it antibonding}} {\rm valence shell (such as} $\pi _g$ or
$\sigma _u$) having a {\rm negligibly small electron density near
the central region between the atomic nuclei in homonuclear
diatomic. Neither the sign {\it ''}}$+${\rm {\it ''}\ always
corresponds to a {\it bonding} valence shell only (such as }
$\sigma _g${\rm \ or }$\pi _u${\rm ) with a considerable electron
density near the internuclear axis region}. {\rm Let us also
recall, that the {\it ''}}$+${\it ''}{\rm \ superposition of
contributing AOs doesn't not necessarily imply only {\it gerade}
(viz. spatially symmetric, like }$\sigma _g$ or $\pi _g${\rm )
molecular valence shells, but it may be also inherent to {\it
ungerade} (viz. antisymmetric}, {\rm like, e.g. }$\pi _u${\rm )
bonding MOs. Accordingly, the {\it ''}}$-${\it ''}{\rm \
superposition may also correspond both to gerade (such as
antibonding }$\pi _g${\rm ) and ungerade (such as bonding }$\pi
_u${\rm ) molecular valence shells. Thus, depending on the spatial
parity of contributing AOs }$\phi _j^{(n)}\left( {\bf r}\right)
${\rm , the ''}$+${\rm ''\ or ''}$-${\rm '' combination\ in the
right-hand side of (18) is chosen to provide a required spatial
and bonding symmetry of composed molecular valence shell.
Meantime, the form of contributing AOs }$\phi _j^{(n)}\left( {\bf
r}\right) $ {\rm in (20)-(23) is generally chosen to adequately
(although, approximately) reproduce the spatial distribution of
electronic density in respective MO\ under consideration relative
to the internuclear axis. In particular, an appropriate choice of
predominantly contributing AOs }$\phi _j^{(n)}\left( {\bf
r}\right) ${\rm \ allows for approximate reproducing a
considerable electron density (inherent to }$\sigma ${\rm \
valence shells) or negligibly small electron density (inherent to
}$\pi ${\rm \ valence shells)} {\rm nearly along the internuclear
axis spatial region, irrespectively of bonding or antibonding
symmetry. For example, the }$3\sigma _g${\rm \ HOMO in }$N_2$ is
well-known as a considerably prolate along the internuclear axis,
so that respective spatial distribution of molecular electron
density can be approximately reproduced by the linear combination
(20) of two predominantly contributing $2p_z$ hydrogen-like atomic
orbitals, which are oriented along the internuclear molecular axis
(supposed to be coincident with spatial $OZ$ axis). Accordingly,
{\rm the }$1\pi _g${\rm \ HOMO in }$O_2$ is well-known to have a
negligible electron density along the internuclear axis and this
feature can be adequately reproduced by a single linear
combination (23) of two predominantly contributing $2p_x$ (or
$2p_y$) scaled hydrogen-like atomic orbitals, which are oriented
perpendicularly to the internuclear axis.

\qquad {\rm We have to mention here that the ''$+$'' combination
of $2p_z$ scaled hydrogen-like atomic orbitals we previously used
in [31] to reproduce} $3\sigma _g$ {\rm HOMO in $N_2$ doesn't
provide the even ({\it gerade}) spatial parity and {\it bonding}
symmetry of the }$\left( 2p\right) 3\sigma _g${\rm , although
proved to be a surprisingly well working in explaining of relevant
experiment [26] (which, particularly, shows no suppression found
in strong-field ionization of $N_2$). Therefore, presently the
''$-$'' combination of $2p_z$ states in Eq.(20) is used, which
does provide the even spatial parity of the }$\left( 2p\right)
3\sigma _g$ {\rm HOMO. However, according to the MO-SFA model
[19], such a composition would result in destructive
intramolecular interference and accordingly high suppression in
ionization of $N_2$ (see also [31], for details). To eliminate the
mentioned deficiency, in our present consideration we apply a more
accurate composition of the }$3\sigma _g${ HOMO taking into
account some admixture of contribution from atomic $s$-states
required mostly to provide a good agreement with experiment (which
doesn't show a suppressed ionization in $N_2$). Thus, the }
$3\sigma _g$ {\rm MO is to be further approximated by the {\it
coherent superposition} of few different MOs corresponding to
separate contributions from atomic states of a specified orbital
symmetry (viz., the scaled hydrogen-like $1s$, $2s$ and $2p_z$
orbitals):}
\begin{equation}
\Phi _{\left( 1s2s2p\right) 3\sigma _g}\left( {\bf r};{\bf
R}_0\right) =A_{1s}\Phi _{\left( 1s\right) 3\sigma _g}\left( {\bf
r};{\bf R}_0\right) +A_{2s}\Phi _{\left( 2s\right) 3\sigma
_g}\left( {\bf r};{\bf R}_0\right) +A_{2p}\Phi _{\left( 2p\right)
3\sigma _g}\left( {\bf r};{\bf R}_0\right) \label{24}
\end{equation}
with the weight coefficients $A_j\ (j=1,2,3)$ being the relative
contributions ($\left| A_j\ \right| \leq 1$) from respective
atomic states and considered as variational parameters to be found
from the equation for minimum of respective molecular binding
energy, the value of which is put to be equal to the experimental
value. Unlike this, the $1\pi _g$ HOMO in $O_2$ is further
supposed to be reproduced by the single $\left( 2p\right) 1\pi _g$
MO (24), which proved to be fairly well working [31] in
interpretation of a high suppression of strong-field ionization
found in relevant experiment [25, 26].

{\rm \qquad The applied velocity gauge (VG) formulation of SFA
consideration of molecular strong-field ionization also implies
analytical representation for Fourier transform }$F_n\left({\bf
q}, {\bf R}_0\right)$ {\rm corresponding to matrix element (6) of
EM transition from the }$n$-th {\rm initial molecular discrete
state }$\Phi_n({\bf r};{\bf R}_0)$ {\rm to the final continuum
state of a definite value of canonical momentum }${\bf p}$. {\rm
For molecular valence shells }$3\sigma _g$ in $N_2$ and $1\pi _g$
in $O_2$ {\rm under consideration, the explicit expressions
(20)-(23) allow for direct analytical calculation of respective
Fourier transform (or molecular wavefunction in momentum space)
for each of contributing one-electron two-centered single
molecular orbitals:}
\begin{equation}
F_{\left( 2p\right) 3\sigma _g}\left( {\bf p}_N{\bf ,R}_0\right)
=- \sqrt {N_e^{(3\sigma _g)}} C\left( \kappa _n\right)
\frac{2^5\kappa _n^{7/2}p_N\cos \left( \theta _{{\bf p} }\right)
}{\pi \sqrt{2}\left( p_N^2+\kappa _n^2\right) ^3}\frac{\sin \left[
\left( {\bf p}_N{\bf \cdot R}_0\right) /2\right] }{\sqrt{2\left[
1-S_{2p_z}\left( R_0\right) \right] }}  \label{25}
\end{equation}
\begin{equation}
F_{\left( 2s\right) 3\sigma _g}\left( {\bf p}_N{\bf ,R}_0\right) =
\sqrt {N_e^{(3\sigma _g)}} C\left( \kappa _n\right)
\frac{2^4\kappa _n^{5/2}\left( p_n^2-\kappa _N^2\right) }{ \pi
\sqrt{2}\left( p_N^2+\kappa _n^2\right) ^3}\frac{\cos \left[
\left( {\bf p}_N{\bf \cdot R}_0\right) /2\right] }{\sqrt{2\left[
1+S_{2s}\left( R_0\right) \right] }}  \label{26}
\end{equation}
\begin{equation}
F_{\left( 1s\right) 3\sigma _g}\left( {\bf p}_N{\bf ,R}_0\right) =
\sqrt {N_e^{(3\sigma _g)}} C\left( \kappa _n\right)
\frac{2^{5/2}\kappa _n^{5/2}}{\pi \left( p_N^2+\kappa _n^2\right)
^2}\frac{\cos \left[ \left( {\bf p}_N{\bf \cdot R}_0\right)
/2\right] }{\sqrt{2\left[ 1+S_{1s}\left( R_0\right) \right] }}
\label{27}
\end{equation}
\begin{equation}
F_{\left( 2p\right) 1\pi _g}\left( {\bf p}_N{\bf ,R}_0\right) =-i
\sqrt {N_e^{(1\pi _g)}} C\left( \kappa _n\right) \frac{2^5\kappa
_n^{7/2}p_N\sin \left( \theta _{{\bf p} }\right) }{\pi
\sqrt{2}\left( p_N^2+\kappa _n^2\right) ^3}\frac{\sin \left[
\left( {\bf p}_N{\bf \cdot R}_0\right) /2\right] }{\sqrt{2\left[
1+S_{1s}\left( R_0\right) \right] }} \label{28}
\end{equation}
{\rm with the polar angle }$\theta _{{\bf p}}$ {\rm of
photoelectron emission relative to internuclear molecular axis
}$\left( \cos \left( \theta _{{\bf p}}\right) =\left( {\bf p}\cdot
{\bf R}_0\right) /pR_0\right) $ {\rm and }$N_e^{(3\sigma _g)}=2$
{\rm and }$N_e^{(1\pi _g)}=4${\rm . Here }$\kappa _n=Z_j^
{(n)}/a_j$ {\rm , whereas }$Z_j^{(n)}$ {\rm is the effective
charge corresponding to ''{\it effective}'' long-range Coulomb
model binding potential of respective residual molecular ion,
while }$ a_j=ja_0${\rm \ is }$j${\rm -th Bohr orbital radius of
respective atomic orbital contributing to }$n${\rm -th initial
molecular discrete state of molecular binding energy }$\varepsilon
_0^{\left( n\right) }=-\kappa _n^2/2=-I_p^{(n)}=-\left(
Z_j^{(n)}/a_j\right) ^2/2$ {\rm under consideration (viz.,
}$\varepsilon _0^{\left( 3\sigma _g\right) }\left( N_2\right)
\approx -15.58$ $eV$, $\varepsilon _0^{\left( 1\pi _g\right)
}\left( O_2\right) \approx -12.13$ $eV${\rm ). The analytical
expressions for respective atomic orbital overlap integrals (19)
are presented in [31], moreover, due to the VG version of SFA
applied, the correction factor }$ C\left( \kappa _n\right) =\left(
2\kappa _nI_p^{(n)}/E\right) ^{\kappa _n^{-1}}${\rm \ is also
introduced to matrix elements (25)-(28) to incorporate the
long-range electron-molecular ion Coulomb interaction in the final
continuum {\it Volkov states} into account [19]}.

\qquad {\rm Let's note first that the angular dependence of the
}$\left( 2p\right) 3\sigma _g$ {\rm molecular wavefunction in
momentum space (25) clearly indicates that a separate contribution
from }$2p_z$ {\rm states alone is highly resistant to ionization
along the direction perpendicular to the internuclear molecular
axis }$\left( \cos \left( \theta _{{\bf p}}\right) \approx
0\right) ${\rm . The respective angular dependence on the angle
}$\theta _{{\bf p}}$ {\rm of photoelectron emission relative to
the internuclaer axis is well seen in Fig.1a to be a considerably
prolate along the internuclear axis. Thus, the photoelectron
emission from }$\left( 2p\right) 3\sigma _g$ {\rm MO is profoundly
dominated along the internuclear molecular axis }$\left( \cos
\left( \theta _{{\bf p}}\right) \approx 1\right) ${\rm . similar
to observed earlier in [31] for incorrect (the }''$+$''{\rm )
combination of two $2p_z$ states. This, particularly, illustrates
that the spatial (gerade or ungerade) symmetry of the }$\left(
2p\right) 3\sigma _g$ {\rm MO doesn't essentially affect the
angular dependence (viz., {\it nodal plane} along the molecular
axis) of respective Fourier transform. Owing to the presence of
$\cos \left( \theta _{{\bf p} }\right)$ factor (see Eq.(25)), both
the ''$-$'' (currently used) and the ''$+$'' (earlier used in
[31]) combination of $2p_z$ atomic states suggest a predominant
photoelectron emission along the internuclear axis (see also
Fig.1a presented in [31], for comparison). The sign of the
combination mostly alters the absolute value of local maxima in
respective Fourier transform due to different ({\it sine} or {\it
cosine}) trigonometric factor arising due to intramolecular
interference of contributions of ionization from two atomic
centers [19] separated by the internuclear distance }$R_0${\rm .
Since the generalized Bessel functions in Eqs.(10)-(11) are
commonly maximized when photoelectron momentum }${\bf p}$ {\rm is
parallel to the laser polarization }${\bf e}${\rm , the ionization
from } $\left( 2p\right) 3\sigma _g$ {\rm HOMO composed as a
single combination of }$2p_z$ {\rm states alone is thus expected
to be predominant when }$N_2$ {\rm molecular axis is aligned along
the laser field polarization (}${\bf R}_{{0}}||{\bf e}${\rm ). The
latter conclusion (also previously made in [31] for }$\left(
2p\right) 3\sigma _g$ {\rm HOMO of ungerade symmetry) seems to be
natural and consistent with relevant experiment [32] and earlier
alternative VG-SFA calculation [19] (see Fig.7b presented therein)
using pure numerical self-consistent Hartree-Fock (HF) procedure
of }$3\sigma_g${\rm HOMO composition taking the contribution from
atomic $s$-states into account. However, this statement is in a
contradiction to respective VG-SFA results of calculation [33]
applying the same VG-SFA approach and numerical HF-based procedure
of }$3\sigma_g$ {\rm composition as proposed earlier in Ref.[19].
Contrary to [19], the strong-field ionization of $N_2$ was found
in [33] to be predominant for perpendicular orientation of
internuclear axis relative to the laser field polarization.}

\qquad {\rm  To clarify the reason of the contradiction, Figures
1b-1d demonstrate the angular dependence of molecular wavefunction
in momentum space calculated for possible different $3\sigma _g$
compositions (5), either with or without taking a comparable
contribution from atomic $s$-states into account. The relative
(weight) coefficients $A_j$ at $1s$ , $2s$ and $2p_z$ states under
consideration proved to have comparable values for $3\sigma _g$
composition presented in Fig.1, although not necessarily of the
same signs. For example, $A_{1s}\simeq -A_{2p}\approx 1/\sqrt{2}$
for $\left( 1s2p\right) 3\sigma _g$, whereas $A_{1s}\simeq
-A_{2s}\simeq -A_{2p}\approx 1/\sqrt{3}$ for $\left( 1s2s2p\right)
3\sigma _g $ composition, so that the weight coefficients at $1s$
and $2p_z$ states (as well as $1s$ and $2s$ states) are found to
have always opposite signs. Moreover, only the combinations with
$A_{1s}$ and $A_{2p}$ (as well as $ A_{1s}$ and $A_{2s}$) of
opposite signs proved to provide no suppression in $ N_2$
ionization revealed in experiment [26]. To summarize, the
presented Fourier transforms corresponding to }$\left(1s2p\right)
3\sigma_g$ and/or $ \left(1s2s2p\right) 3\sigma_g$ {\rm valence
shells clearly suggest a predominant photoelectron emission from
$3\sigma _g$ HOMO along the internuclear axis, contrary to
suggested in [33]. Our Fig.1 also demonstrates that such
difference may only arise owing to relative contribution from
}$s$-states, which seems to be somehow {\it overestimated}
(compared to $p$-states) within the particular (12s7p)/[6s4p]
basis chosen in [33] for numerical HF-based procedure of
$3\sigma_g$ composition in coordinate space (see also [34] for
details). In the meantime, a composition overestimating the
relative contribution from $s$-states doesn't seem to be suitable
to $3\sigma _g$ molecular state, which is known to have a more
complex structure surely {\it dominated} by $p$-states in
coordinate space [17]. This also becomes especially evident owing
to Fig.12 in Ref.[35] displaying the correct orbital shape for the
$3\sigma _g$ wavefunction in $N_2$, which obviously contrasts with
the resulting $3\sigma _g$ state of substantially different shape
and detailed structure suggested in [33] (cf., for example,
respective Fig.1c presented in the last reference therein).

\qquad {\rm For laser-irradiated atomic counterparts of diatomics
under consideration - such as }$Ar$ {\rm (for }$N_2$ {\rm and/or
}$H_2${\rm ) and }$Xe$ {\rm (for} $O_2${\rm ) with }$3p$ {\rm and
}$5p$ {\rm outermost atomic valence shells, respectively - the
initial atomic ground state can be also approximately reproduced
by corresponding one-electron hydrogen-like atomic orbital. The
analytical expression for respective atomic Fourier transforms
}$F_{3p_z}^{\left( 3p_x,3p_y\right) }\left( {\bf p}\right) ${\rm \
of }$3p_z${\rm , }$3p_x$ {\rm and }$3p_y$ {\rm hydrogen-like AOs
(e.g., in }$Ar${\rm ) corresponding to different possible values
of associated magnetic quantum number }$m$ {\rm (}$0$ {\rm \ - for
}$3p_z${\rm \ and }$\pm 1${\rm \ - for }$3p_x\pm i3p_y${\rm ) has
the form:}
\begin{equation}
F_{3p_z}^{\left( 3p_x\pm i3p_y\right) }\left( {\bf p}_N\right)
=\frac{ iC\left( \kappa _n\right) 2^4\sqrt{2}\kappa
_n^{11/2}p_N}{\pi \sqrt{3}\left( p_N^2+\kappa _n^2\right)
^4}\left(
\begin{array}{c}
\pm \sin \left( \theta \right) \exp \left( \pm i\varphi \right)  \\
\sqrt{2}\cos \left( \theta \right)
\end{array}
\right)   \label{29}
\end{equation}
{\rm Here the angles }$\theta ${\rm \ and }$\varphi $ {\rm are the
polar and azimuthal angles of photoelectron emission with respect
to the incident field polarization (}$\cos \left( \theta \right)
=\left( {\bf e}\cdot {\bf p} \right) /p${\rm ), so that the
angular factor in the right-hand side of (29) corresponds to the
case of either }$3p_x\pm i3p_y$ ({\rm the upper part), or
}$3p_z${\rm \ (the lower part) AO. The ''effective'' charge
}$Z_{eff}^{(n)}$ {\rm of pure Coulomb model binding potential
corresponds to the experimental value of atomic ionization
potential }$I_p=\kappa _n^2/2=\left( Z_{eff}^{(n)}/a_3\right)
^2/2$.{\rm \ (}$I_p\approx 15.75$ $eV${\rm \ - for } $Ar${\rm \
and }$I_p\approx 12.13$ $eV${\rm \ - for }$Xe${\rm ).}

\qquad {\rm The proposed strong-field model is also equally well
suitable to describe molecular HHG process in arbitrary molecular
valence shell including inner ones closest to the HOMO (although,
of higher binding energy, different orbital symmetry and number of
identical electrons). The contributions from different (outermost
and inner) valence shells are to be added {\it coherently} in
molecular HHG, so that total amplitude of emission of }$N${\rm -th
harmonic is just the sum of respective partial amplitudes
corresponding to contribution from each molecular valence shell
under consideration. In particular, the overall }$N${\rm -th
partial amplitude (4) of harmonic emission in diatomic }$N_2${\rm
\ can be represented as consisting of contributions from three
highest molecular valence shells - }$ 3\sigma _g$ ($I_p^{\left(
3\sigma _g\right) }\approx 15.58$ $eV$), $1\pi _u$ ($I_p^{\left(
1\pi _u\right) }\approx 16.96$ $eV$) and $2\sigma _u$ ($
I_p^{\left( 2\sigma _u\right) }\approx 18.73$ $eV$){\rm :}
\begin{equation}
f_N^{\left( N_2\right) }\left( \Omega _{{\bf k}^{\prime}}\right)
=f_N^{\left( 3\sigma _g\right) }\left( \Omega _{{\bf k}^{\prime
}}\right) +f_N^{\left( 1\pi _u\right) }\left( \Omega _{{\bf
k}^{\prime }}\right) +f_N^{\left( 2\sigma _u\right) }\left( \Omega
_{{\bf k}^{\prime }}\right) \label{30}
\end{equation}
{\rm whereas, for HHG in diatomic }$O_2${\rm , accordingly, as
contribution from three different highest molecular valence shells
- }$1\pi _g$ ($ I_p^{\left( 1\pi _g\right) }\approx 12.07$ $eV$),
$1\pi _u$ ($I_p^{\left( 1\pi _u\right) }\approx 16.26$ $eV$) and
$3\sigma _g$ ($I_p^{\left( 3\sigma _g\right) }\approx 18.18$
$eV$):
\begin{equation}
f_N^{\left( O_2\right) }\left( \Omega _{{\bf k}^{\prime }}\right)
=f_N^{\left( 1\pi _g\right) }\left( \Omega _{{\bf k}^{\prime
}}\right) +f_N^{\left( 1\pi _u\right) }\left( \Omega _{{\bf
k}^{\prime }}\right) +f_N^{\left( 3\sigma _g\right) }\left( \Omega
_{{\bf k}^{\prime }}\right) \label{31}
\end{equation}
\qquad {\rm Under the conditions of equivalent atomic centers and
the atomic or molecular specie occupies the same state before and
after the passage of the laser pulse, only odd harmonics of the
fundamental laser frequency of linear polarization are emitted
[29]. The harmonic emission rates are to be currently calculated
only for high-order harmonics of polarization ${\bf e}_{{\bf
k}^{\prime },\lambda ^{\prime }}$, for which the emission rate
becomes maximal, that is parallel to polarization ${\bf e}$ of
incident laser field. The model-related background analytical
expressions (3)-(12) and (25)-(28) do imply fully $3D$
consideration of the process (particularly, they are available for
arbitrary orientation of internuclear molecular axis with respect
to the laser polarization ${\bf e}$), nonetheless, the molecular
axis is supposed further to be strongly aligned along the
polarization of incident laser field (i.e. $\left({\bf e}\cdot
{\bf R}_0\right) \sim \cos \left( \Theta \right) =0$ ) throughout
our present consideration, unless stated otherwise. The associated
{\it orientation effects} in molecular harmonic generation are
thus beyond the scope of our present consideration, which is
mostly focused on behavior of molecular harmonic spectra depending
on equally important problem parameters, such as internuclear
separation $R_0$ and properties of a specified laser-irradiated
diatomic (such as orbital symmetry, composition and structure of
molecular valence shells) as well as incident laser parameters
(such as laser intensity, pulse duration and temporal profile).}\\

\item  CALCULATION, NUMERICAL RESULTS AND DISCUSSION.

\qquad {\rm When studying the strong-field phenomena, the major
difficulty with molecules is to deal with both the electronic and
the nuclear degrees of freedom (i.e., related to the molecular
vibrational and rotational motion). Nonetheless, the including of
the nuclei motion into a proper consideration is generally
feasible within frame of the currently applied approach, but this
additionally requires of preliminary knowing available reliable
data (previously calculated or measured in experiment) related to
dependence of molecular binding energy on internuclear separation
$R_0$. To the best of our knowledge, the latter dependence is
reliably known only for the simplest diatomics ($H_2^{+}$ or
$H_2$), solely for which the $R_0$ -dependent behavior of
molecular HHG can be numerically studied within the currently
proposed model. For all other diatomics under consideration ($N_2$
and $O_2$), the nuclei motion will be further completely ignored
and the {\it Born-Oppenheimer approximation} is supposed to be
valid. For this reason we start our numerical study from
$H_2^{+}$\ molecular ion, which is sufficiently well explored in
previous theoretical and experimental studies (see [36], for a
review). Beginning from the first HHG calculations [37, 38], this
molecular specie, owing to its structural simplicity, became the
object of a large number of other theoretical and experimental
investigations (e.g., [8, 10-12]), the results of which can be
used for direct comparison.}

\qquad {\rm Fig.2 displays our results for high-harmonic spectra
of $H_2^{+}$ \ calculated as harmonic emission rates (13) compared
to respective atomic counterparts of nearly identical binding
energy under conditions considered earlier in [10]. Like in [10],
the results presented in Fig.2a were calculated assuming
$R_0$-independent fixed molecular binding energy $ \varepsilon _0$
(viz., $\left| \varepsilon _0\right| =0.4$ $a.u.$ for any
internuclear separation $R_0$), which is supposed to be identical
to that of model atomic ''counterpart'' bound to model zero-range
potential (ZRP). Since the molecular binding energy is known to be
$R_0$-dependent, Fig.2b also presents the analogous high-harmonic
spectra of }$H_2^{+}${\rm \ calculated under the same laser pulse
and $R_0$-dependent molecular binding energy $\varepsilon _0\left(
R_0\right) $ corresponding to initial $1\sigma _g$ ground state.
Also, unlike Fig.2a, the high-harmonic spectrum of respective
atomic ''counterpart'' was calculated assuming the hydrogen-like
$1S$ discrete state of atomic binding energy $\varepsilon
_0\approx -15.86\ eV$ nearly identical to the molecular binding
energy $\varepsilon _0\approx -0.583\ a.u.$ of }$H_2^{+}\left(
1\sigma _g\right) ${\rm \ corresponding to $ R_0=12$ $a.u.$ The
molecular harmonic spectra presented in Fig.2 show the clear signs
of the so-called ''{\it low-energy hump}'' similar to that
revealed in [10] within a low-frequency domain, which is
enormously enhanced for two-centered case as compared to the
one-centered (''atomic'') case also presented in Fig.2. The origin
of this ''hump'' arising in molecular high-harmonic spectra was
well explained earlier [10] in terms of the {\it rescattering}
mechanism and, particularly, by the existence of two different
pathways for the initially released electron to recombine: with
the center it started from as well as with the other. Namely, the
''hump'' corresponds to the case when the initially released
electron moves directly from one center to the other within a time
interval that is short compared with the period of the field,
moreover, this happens around the time where the field is near its
maximum. Under these conditions the energy gain, which the
released electron acquires from the driving laser field when
moving along the shortest way from one atomic center to other, can
be approximately evaluated just by $\varepsilon _{el}\sim ER_0$
that corresponds the position of the top harmonic $N_h\approx
ER_0/\omega $ of the ''hump''. According to this quasiclassical
interpretation, the position of its top harmonic should be
noticeably shifted toward a higher frequency part for a larger
internuclear separation $R_0$. This particular behavior is fairly
well reproduced in Fig.3 by other high-harmonic spectra of
$H_2^{+}$ calculated for a different incident laser pulse and
various larger values of $R_0$, for which Fig.3 also shows the
signs of ''split'' arising within the ''low-energy hump''.}

\qquad {\rm We also note the position of high-frequency plateau
cutoff calculated for relatively large internuclear separation and
presented in Figures 2 and 3 noticeably deviates from what is
prescribed by the well known rule ''$I_p+3.17U_p$'' for standard
harmonic cutoff frequency. So, according to this rule, one would
expect for the position }$N_C${\rm \ of the cutoff frequency to be
nearly at the harmonic order $N_C\approx 43$ in Fig.2a or
$N_C\approx 49$ in Fig.2b, that is quite consistent with those
figures suggest for atomic case. Meanwhile, the respective
molecular harmonic spectra calculated for $R_0=12\ a.u.$ clearly
demonstrate a noticeably different position for the cutoff
frequency, viz., at harmonic order $N_C\approx 51$ in Fig.2a or
$N_C\approx 55$ in Fig.2b. Moreover, there is also a pronounced
and well-defined ''{\it cutoff split}'' visible within the plateau
cutoff domain of molecular harmonic spectra presented in Figs.2,
3. This ''cutoff split'' is similar to the split revealed earlier
[10] in the cutoff domain of $H_2^{+}$ high-harmonic spectra
calculated for sufficiently large internuclear separation $R_0$.
The origin of the ''cutoff split'' can be also well explained in
terms of the rescattering mechanism and, particularly, attributed
to the interference of two other (longest, but topologically
different) pathways for the released electron moving from the one
atomic center to the other to recombine with (see, also [10], for
details). The respective separation in energy the released
electron gains from the driving laser field is about at the
position of high-frequency plateau cutoff (viz., nearly around the
mentioned above harmonic order $N=47$ in Fig.2a and/or $N=51$ in
Fig.2b), so that the two maxima alongside completely mask the
actual cutoff position as the distance $R_0$ increases.}

\qquad {\rm As noted above, unlike the atomic case, the cutoff
position $N_C$ in calculated molecular harmonic spectra proved to
be not fixed yet even for the fixed parameters of incident laser
pulse, but also substantially dependent on internuclear
separation. Such }$R_0$-{\rm dependent behavior of $N_C\left(
R_0\right) $ is well illustrated by Fig.3, which represents $
H_2^{+}$ harmonic spectra calculated for different fixed values of
$R_0$ under the same laser pulse. In particular, Fig.3
demonstrates the deviation of the harmonic cutoff position from
the well known rule ''$I_p+3.17U_p$'', even if the corresponding
decrease of molecular ionization potential $I_p$ for larger $R_0$
is taken into complete account. So, in Fig.3a the position of
high-frequency cutoff for relatively small $R_0=4\ a.u.$ is in a
full accordance with the aforementioned\ rule for standard cutoff
corresponding to the harmonic order $N_C\approx 73$. This value is
accordingly and noticeably smaller compared to $N_C\approx 80$
prescribed for larger ionization potential $I_p=30.0\ eV$
corresponding to the equilibrium internuclear separation $R_0=2\
a.u.$ Thus, the calculated position of harmonic cutoff initially
moves toward a lower frequency domain of spectrum due to
corresponding decrease of molecular ionization potential $I_p$ as
$R_0 $ increases. However, under further expansion and
corresponding decrease of molecular ionization potential, the
cutoff frequency is, nevertheless, getting to move back toward a
higher frequency domain, contrary to the ''standard cutoff'' rule.
So, for example, in molecular harmonic spectra calculated for
larger internuclear separation and presented in Fig.3a, one can
identify the cutoff position located at harmonic order $N_C\left(
R_0=15\ a.u.\right) \approx 79$ and $N_C\left( R_0=30\ a.u.\right)
\approx 89 $ respectively. These two values are noticeably larger
than $N_C\approx 70$ prescribed by the mentioned ''standard
cutoff'' rule based on correspondingly decreased molecular
ionization potential in $H_2^{+}$ taken into account. The physical
mechanism underlying this quite an opposite tendency of increasing
the harmonic cutoff frequency for sufficiently large $ R_0$
increased can be again well understood in terms of semiclassical
''recollision'' picture invoked above for interpretation of the
''cutoff split''. Namely, when $R_0$ increases, the released
electron acquires an additional energy gain from the driving laser
field when moving along the longer ways from one atomic center to
other. Due to the mentioned additional energy gain followed by
recombination the acquired electron energy is released as a photon
of a considerably higher frequency that gives rise to increasing
of the harmonic cutoff frequency. Beginning from sufficiently
large internuclear separation (over about $R_0=10\ a.u.$), this
mechanism becomes predominant, insomuch that the related
additional energy gain can compensate or even overcome the
competing decrease of standard cutoff frequency due to diminished
molecular ionization potential $I_p$ for larger $ R_0$. Such a
complicated $R_0$-dependent behavior of harmonic cutoff position
revealed in molecular high-harmonic spectra under discussion seems
to be very promising for various applications. To the best of our
knowledge, the dissociating diatomics with large internuclear
separations were predicted first in [39] to be capable to produce
high harmonics of frequency far beyond the ''standard'' (atomic)
cutoff, however, this idea has not been confirmed yet by
experiments. Particularly, when the nuclei are separated by a
distance equal to odd multiples of $\pi \alpha _0$ (where $ \alpha
_0=\left( E/\omega ^2\right) $ is the classical amplitude of
electron's excursion in laser field), the cutoff limit was shown
to extend beyond $I_p+3.17U_p$ up to $I_p+8U_p$ since the limit of
$8U_p$ is the maximum kinetic energy reachable by the electron
during its classical excursion from one atomic center to other
(see also [10]). The results presented in Fig.3 generally confirm
such a prediction and indicate unambiguously that, quite a
noticeable {\it extension} of high-frequency plateau cutoff is to
be present in molecular harmonic spectra for very large
internuclear separation. Although, due to relatively small values
of $R_0$ used under calculations compared to $\pi \alpha _0$
$\approx 110\ a.u.$, the calculated extension of high-frequency
plateau is not very impressive. }

\qquad {\rm On the other hand, the values of $R_0$ considered in
Figures 2, 3 proved to be sufficiently large to confirm the result
of previous research [10, 38] showing that molecular harmonic
spectra do not obey the ''$ I_p+3.17U_p$'' rule yet for the
position of high-frequency cutoff. Moreover, unlike atomic
high-harmonic spectra, the molecular harmonic cutoff is split due
to intramolecular interference and this ''cutoff split'' well seen
to be a noticeably broader and more pronounced for larger $R_0$.
Also, like in [10], the currently calculated $H_2^{+}$ harmonic
spectra in Fig.3 clearly demonstrate a significant enhancement (up
to several orders of magnitude) in harmonic emission rate as
internuclear separation $R_0$ increases. This is not surprising
for diatomics since the respective molecular binding energy $
\left| \varepsilon _0\left( R_0\right) \right| $ is to be smaller
for larger $R_0$. The similar enhancement of harmonic emission due
to decreased ionization potential was initially revealed in atomic
high-harmonic spectra (see, e.g., Fig.5 and Fig.10 in [40] as well
as references to relevant experiments cited therein). This also
allows for quite a clear explanation, namely, the laser-irradiated
species having a smaller ionization potential are generally more
polarizable, so that their response to the incident laser field is
expected to be a noticeably stronger than in harder ionized
species. Our results calculated for $H_2^{+}$ and presented in
Fig.3 also confirm that the similar underlying physical mechanism
is responsible for the enhancement of harmonic emission rate in
diatomics for larger internuclear separation as well as the
transient enhancement of the high-harmonic conversion efficiency
recently found as exceeding the atomic one in expanding molecules
[12, 14]. It is also very interesting that Fig.3 still
demonstrates noticeable differences in details of two
high-harmonic spectra of $H_2^{+}$ corresponding to different
$1\sigma _g$ and $1\sigma _u$ initial molecular states and
calculated for large $R_0=30\ a.u.$, at which generic structural
features (e.g., such as orbital symmetry) of particular
laser-irradiated diatomic are expected to have a very minor
importance. }

\qquad {\rm Another remarkable feature well recognized in
molecular harmonic spectra presented in Fig.3a is a sufficiently
pronounced local minimum clearly visible within the high-frequency
plateau region and located respectively at }$N=55$ for {\rm
$R_0=4\ a.u.$, }$N=53$ for {\rm $R_0=15\ a.u.$ and }$N=49$ for
{\rm $R_0=30\ a.u.$ The minima seem to be similar to those
revealed earlier in numerically calculated molecular high-harmonic
spectra [11] and predicted to arise in harmonic spectra of
diatomics with HOMO of bonding symmetry due to {\it destructive}
intramolecular interference of photoelectron emission from
separated atomic centers to intermediate continuum states. Like in
[11], the interference-related minimum in Fig.3a shows a regular
dependence of its position on internuclear separation and,
particularly, moves toward a lower frequency domain with increase
of $R_0$. Such $R_0$-dependent behavior is generally in accordance
with the semiclassical (although, quite an {\it ad hoc})
interpretation given in [11] and related rule }$N_{\min }^{(HHG)}=
I_p/\omega +\pi ^2/\left( 2\omega R_0^2\right) \sim \pi ^2/\left(
2\omega R_0^2\right) $ {\rm suggested for location of the minimum.
However, the harmonic spectra calculated under the conditions of
[11] and presented in Fig.3a doesn't seem to confirm the above
prediction for location of the minimum, but rather suggest a
different (although, quite a raw) approximate rule }$N_{\min
}^{(HHG)}= {(I_p+U_p)}/\omega +\pi ^2/\left( 2\omega R_0^2\right)$
{\rm . The presently suggested empirical rule seems to be similar
to revealed earlier for the interference-related minimum arising
in calculated molecular photoelectron spectra [10, 31] due to
direct ATI in diatomics with HOMO of bonding symmetry. Therefore,
the interference-related phenomena arising in molecular harmonic
spectra should also allow for transparent interpretation similar
to that given for molecular ATI [31]. In our opinion, this also
remarkably demonstrates a close and intimate interrelation of
molecular ATI and HHG processes.}

\qquad {\rm Since the harmonic generation process is strongly
related to ionization, one should also expect the enhanced
harmonic emission rates in laser-irradiated species for which the
ionization rate is also enhanced for some reason. On the other
hand, for laser pulses of sufficiently high peak intensity
}$I${\rm \ and long pulse duration $\tau $, the enhanced
ionization leads to a faster saturation of ionization and
exhaustion of initial state of laser-irradiated system, which,
after reaching the saturation of ionization.is unable to
efficiently produce high-order harmonics anymore. This gives rise
to quite a natural restriction imposed on standard cutoff
frequency $I_p+3.17U_p$ (with ponderomotive energy $U_p\sim I $),
which thus cannot be extended yet under further increase of
incident laser intensity $I$ over the saturation intensity
$I_{sat}$. Hence, the laser-irradiated species, such as $H_2$ and
$O_2$, showing a suppression in ionization rate (compared to
counterpart systems of nearly identical ionization potential
$I_p$) are expected to produce harmonics of considerably higher
frequency corresponding to a longer high-frequency plateau with an
extended cutoff due to a noticeably higher laser intensity }$
I_{sat}${\rm \ that saturates ionization. The extension in
high-harmonic spectra of $O_2$ relative to atomic $Xe$ under
respective laser saturation intensity was recently observed in
experiment [13] and also partly reproduced in related MO-ADK
calculations [13] based on the Lewenstein atomic HHG\ model [22].}

{\rm \qquad In order to assess a possible extension of
high-frequency plateau in molecular harmonic spectra of more
complex diatomics, such as $N_2 $ and $O_2$, compared to
respective atomic counterparts, one has to evaluate first the
respective saturation laser intensity for each of laser-irradiated
species under consideration. Fig.4 displays our results for total
probability of ionization of $O_2$, $Xe$, $N_2$ and $Ar$
calculated under conditions of experiment [13] according to
formula (16) extended to the diatomic case based on earlier
developed SFA-LCAO model [31] as described in Sec.3 above. The
respective saturation laser intensity $I_{sat}$ is well identified
in Fig.4.and defined as the intensity at which about $99\%$ of
laser-irradiated species being ionized by the end of the laser
pulse duration. The results have been summarized in the Table I,
which, for the convenience of direct comparison, also represents
the respective results of [13] calculated according to MO-ADK
theory [23].}

\begin{center}
{\bf Table I}\\[10pt]

\begin{tabular}{|c|c|c|c|c|}
\hline
Specie & $N_2$ & $Ar$ & $O_2$ & $Xe$ \\ \hline
$
\begin{array}{c}
{\rm HOMO} \\
{\rm AO}
\end{array}
$ & $
\begin{array}{c}
\left( 1s2s2p\right) 3\sigma _g
\end{array}
$ & $
\begin{array}{c}
\\
3p^6
\end{array}
$ & $
\begin{array}{c}
\left( 2p\right) 1\pi _g
\end{array}
$ & $
\begin{array}{c}
\\
5p^6
\end{array}
$ \\ \hline
$R_0\left( \AA \right) $ & $1.1$ &  & $1.21$ &  \\ \hline
$I_p\left( eV\right) $ & $15.58$ & $15.75$ & $12.07$ & $12.13$ \\ \hline
$
\begin{array}{c}
I_{sat}\left( W/cm^2\right)  \\
{\rm present} \\
{\rm calculation}
\end{array}
$ & $3.75\cdot 10^{14}$ & $4.4\cdot 10^{14}$ & $2.5\cdot 10^{14}$
& $ 1.7\cdot 10^{14}$ \\ \hline $
\begin{array}{c}
I_{sat}\left( W/cm^2\right)  \\
{\rm \ MO-ADK } \\
{\rm calculation [13]}
\end{array}
$ & $3.71\cdot 10^{14}$ & $4.4\cdot 10^{14}$ & $3.01\cdot 10^{14}$
& $ 1.75\cdot 10^{14}$ \\ \hline
\end{tabular}
\end{center}

{\rm The table demonstrates a fairly good consistence of our
currently calculated results with the prediction of MO-ADK theory,
except the saturation intensity calculated for }$O_2${\rm . The
deviation is explained by a different value of molecular
ionization potential, viz. $I_p\left( O_2\right) =12.56\ eV$ used
in [13] (see, e.g., Table I presented therein), which is a
slightly higher compared to the currently used experimental value
$I_p\left( O_2\right) =12.07\ eV$ [26]. This inevitably leads to
an overestimated high suppression in ionization of $O_2$ relative
to $Xe$, thereby giving rise to a noticeably overestimated value
of the saturation laser intensity $3.01\cdot 10^{14}W/cm^2$ as
compared to $2.5\cdot 10^{14}W/cm^2$ currently calculated.}

\qquad {\rm The high-harmonic spectra of $O_2$ presented in Fig.5
demonstrate a noticeable extension compared to the spectrum of
$Xe$ calculated at respectively lower laser saturation intensity
previously found (Fig.5b). At the same time, the extension of
molecular high-frequency plateau and respective cutoff frequency
currently calculated for $O_2$ (relative to $Xe$) noticeably
deviates from the similar extension found in [13] for different
saturation intensities calculated using the MO-ADK model.
Particularly, the harmonic order $N_C$ for cutoff position in
presently calculated harmonic spectrum of $O_2$ is about $39$
(versus calculated $43$ and measured $53$, both reported in [13])
provided the contribution from the $1\pi _g$ HOMO is only taken
into account, like in [13]. However, the total harmonic spectrum
of $O_2$ presented in Fig.5a and corresponding to the coherent
contribution (31) from the $\left( 2p\right) 1\pi _g$ HOMO and two
closest $\left( 2p\right) 1\pi _u$ and }$\left( 1s2s2p\right)
3\sigma _g$ {\rm \ inner shells additionally taken into account
demonstrate a noticeably longer high-frequency plateau of cutoff
located at $N_C\left( O_2\right) =45$ and thus extended relative
to $Xe$, for which the cutoff is located at $ N_C\left( Xe\right)
=29$. Note also that the additional extension clearly visible in
Fig.5a is mostly due to the contribution from the inner $3\sigma
_g$ valence shell, whose separate contribution within the cutoff
domain is comparable with that from the $\left( 2p\right) 1\pi _g$
HOMO and always prevailing with respect to that from the $\left(
2p\right) 1\pi _u$, despite the inner $3\sigma _g$ has the higher
ionization potential than $1\pi _g$ or $1\pi _u$. This noted
feature seems to be quite a natural for strongly aligned diatomics
under consideration and can be explained by an enhanced ionization
(initial release of electron to intermediate continuum states)
from the $3\sigma _g$ MO, which is a considerably prolate along
the internuclear axis. Unlike $\sigma $ orbitals, $\pi $ molecular
orbitals (such as $1\pi _g$ HOMO and inner $1\pi _u$ in }$O_2${\rm
) are predominantly oriented along the direction perpendicular to
the internuclear axis, so that the photoelectron emission (initial
release of electron to intermediate continuum states) from $\pi $
molecular orbitals in aligned diatomics is a considerably
suppressed along the internuclear axis that leads to a relatively
smaller harmonic emission rate (see also the relevant discussion
in [31]).}

\qquad {\rm The high-harmonic spectra of $N_2$ and its atomic
counterpart $Ar $ calculated for respective saturation intensities
currently found and shown in Table I, are presented in Fig.6. In
particular, Fig.6b demonstrates some minor extension of cutoff in
harmonic spectrum of $Ar$ relative to $N_2$ due to a slightly
enhanced ionization rate of $N_2$ observed in experiments [25, 26]
and related lower laser saturation intensity compared to $Ar$. The
cutoff position in currently calculated harmonic spectrum of $N_2$
corresponding to the separate contribution from the }$\left(
1s2s2p\right) 3\sigma _g${\rm \ HOMO (only taken into account in
[13]) is located at harmonic order about $55$ that is well
consistent with that calculated in [13]. Meanwhile, the total
harmonic spectrum of $N_2$ presented in Fig.6a and calculated
taking the coherent contribution (30)from two closest inner
$\left( 2p\right) 1\pi _u$ and $\left( 2s\right) 2\sigma _u$
valence shells into account demonstrates a bit longer plateau of
cutoff position at harmonic order $N_C\left( N_2\right) =61$ that
is in a reasonably good accordance with experimental value
measured in [13]. This noted extension of high-frequency plateau
cutoff is also clearly visible to arise mostly due to $2\sigma _u$
valence shell, whose separate contribution within the cutoff
region is comparable with that from the $3\sigma _g$ HOMO and
always prevailing over that from $1\pi _u$, even though the
ionization potential of $2\sigma _u$ is higher. Thus, the
additional extension due to the contribution from inner valence
shells is quite similar to that we observed earlier in the total
high-harmonic spectrum of $O_2$ presented in Fig.5a. This probably
might also explain the difference between the positions found for
harmonic cutoff in high-harmonic spectra of $N_2$ calculated and
observed in [13].}

\qquad {\rm All the results presented in Fig.5 and Fig.6 have been
summarized in Table II below, for the convenience of direct
comparison with experiment [13].}

\begin{center}
{\bf Table II}\\[10pt]

\begin{tabular}{|c|c|c|c|c|}
\hline
{\rm Specie (AO, MO}) & $I_p\left( eV\right) $ & $
\begin{array}{c}
N_C \\
{\rm presently} \\
{\rm calculated}
\end{array}
$ & $
\begin{array}{c}
N_C \\
{\rm MO-ADK\ based} \\
{\rm calculation\ [13]}
\end{array}
$ & $
\begin{array}{c}
N_C \\
{\rm measured\ in} \\
{\rm experiment\ [13]}
\end{array}
$ \\ \hline $N_2\left[ \left( 1s2s2p\right) 3\sigma _g\right] $ &
$15.58$ & $55$ & $55$ &  \\ \hline $N_2\left[ \left( 2p\right)
1\pi _u\right] $ & $16.96$ & $57$ &  &  \\ \hline
$N_2\left[ \left( 2s\right) 2\sigma _u\right] $ & $18.73$ & $57$ &  &  \\
\hline
$N_2\left( 3\sigma _g+1\pi _u+2\sigma _u\right) $ &  & $61$ &  & $57$ \\
\hline
$Ar\left( 3p^6\right) $ & $15.75$ & $67$ & $65$ & $63$ \\ \hline
&  &  &  &  \\ \hline
$O_2\left[ \left( 2p\right) 1\pi _g\right] $ & $12.07$ & $39$ & $45$ &  \\
\hline $O_2\left[ \left( 2p\right) 1\pi _u\right] $ & $16.26$ &
$43$ &  &  \\ \hline $O_2\left[ \left( 1s2s2p\right) 3\sigma
_g\right] $ & $18.18$ & $43$ &  &
\\ \hline
$O_2\left( 1\pi _g+1\pi _u+3\sigma _g\right) $ &  & $45$ &  & $53$
\\ \hline $Xe\left( 5p^6\right) $ & $12.13$ & $29$ & $29$ & $29$ \\
\hline
\end{tabular}
\end{center}

{\rm As a final remark to the results presented in Table II, we
have to comment the respective HHG results for $N_2$\ and $Ar$
reported in [13] and based on MO-ADK calculation of laser
saturation intensity (see Table I in Ref.[13]). Namely,the latter
results seem to be at least confusing and contradicting to the
MO-ADK results presented {\it ibidem} in Fig.4, which
unambiguously shows a considerable suppression for $N_2^{+}$ ion
signal compared to $Ar^{+}$ for the shorter laser pulse of
duration $\tau =30\ fs$. Under these conditions close to
experiment [13] corresponding to $\tau =25\ fs$, the MO-ADK model
suggests the calculated ratio $N_2^{+}/Ar^{+}$ for ion signals
from $N_2^{+}$ relative to $Ar^{+}$ to be nearly around $0.5 $
(see also Fig.8 in [23]) that is consistent with relevant
experiment [25], at least within the strong-field domain where the
tunneling ionization theory is only valid. This, however, implies
that the MO-ADK based calculation would result in an accordingly
higher saturation intensity rather for $N_2$ than for $Ar$,
similarly to the suppressed ionization in $O_2$ relative to $ Xe$.
Contrary to such a natural conclusion the calculated laser
saturation intensity for $N_2$ was identified in [13] to be a
noticeably lower (viz., $ 3.71\cdot 10^{14}$ $W/cm^2$ versus
$4.4\cdot 10^{14}$ $W/cm^2$ reported for $ Ar$) that seems to be
self-contradicting and thus makes a fairly good agreement of the
calculated and measured harmonic cutoff frequency of respective
high-harmonic spectra observed in [13] quite a questionable and
controversial.}\\

\item  CONCLUSION.

\qquad {\rm The strong-field process of high-order harmonic
generation was considered and studied numerically in a number of
laser-irradiated homonuclear diatomics (namely, $H_2^{+}$, $N_2$
and $O_2$) compared to their atomic counterparts of nearly
identical ionization potential. The proposed strong-field
molecular HHG model is essentially based on quantum-mechanical
SFA-based approach developed earlier and currently extended to
molecular case using the standard LCAO-MO method applied for
adequate reproducing the wavefunction of initial molecular
discrete state. The latter wavefunction is approximately
considered as a {\it single-electron} and {\it two-centered}
molecular orbital (MO) consisting of coherent superposition of few
predominantly contributing scaled hydrogen-like atomic orbitals
(AO) centered on each of the atomic cores and thus separated by
internuclear distance $R_0$. The form of contributing AOs is
generally different for various molecular valence shells (such as
$\sigma _g$ , $\pi _u$ and $\pi _g$) of different orbital and
bonding symmetry. Wherever appropriate, the comparable
contributions from other (inner) molecular valence shells of a
larger binding energy and different orbital symmetry are also
incorporated. This, particularly, allows for fully $3D$
consideration of molecular HHG process as well as deriving the
final expressions for amplitude and rate of harmonic emission rate
in closed analytical form available for further numerical
calculations.}

\qquad {\rm The calculated molecular high-harmonic spectra were
found to exhibit a general shape similar to atomic ones, with
extended high-frequency plateau ending by a well defined harmonic
cutoff. Although, the detailed structure, such as the extent of
high-frequency plateau and respective harmonic emission rates
demonstrate a number of noticeable and remarkable differences from
respective atomic harmonic spectra. These distinct differences,
such as the ''{\it low-frequency hump}'' and ''{\it cutoff
split}'' arising due to {\it two-centered} nature of molecular
bound state, are ascertained to be strongly dependent on the
internuclear separation $R_0$. For sufficiently large $R_0$ the
molecular harmonic spectra demonstrate a plateau of a noticeably
longer extent and significant enhancement in harmonic emission
rates compared to spectra corresponding to equilibrium
internuclear separation and calculated under the same laser
pulses. In addition, our SFA calculations seem to confirm the
existence of a broad and pronounced minimum predicted to arise in
harmonic spectra of diatomics with HOMO of bonding symmetry due to
{\it destructive} intramolecular interference of contribution from
photoelectrons of large momenta emitted from separate atomic
centers [11] to intermediate continuum states. Namely, our results
presented in Fig.3a seem to reproduce $R_0$-dependent behavior of
the interference-related minimum found in }$1D${\rm \ and/or
}$2D${\rm \ TDSE calculations [11], although also suggest a bit
different (compared to [11]) approximate rule for location of this
minimum in molecular harmonic spectra.}

\qquad {\rm Finally, the average harmonic emission rate proved to
be very sensitive to orbital symmetry of molecular orbital and its
predominant orientation with respect to the internuclear molecular
axis, insomuch that for some group of harmonics the contribution
from inner molecular shell of higher binding energy (but,
different orbital symmetry) can be comparable (or even predominant
within the plateau cutoff region) relative to that from the HOMO.
Also, due to a high suppression revealed in ionization rate for
$O_2$ relative to $Xe$ atom, the harmonic spectrum calculated for
$O_2$ was found to demonstrate a high-frequency plateau of a
considerably longer extent (and accordingly higher harmonic cutoff
frequency) than in spectrum of $Xe$ atom, for which the saturation
laser intensity was calculated to be a noticeably smaller. To
conclude, the numerically studied behavior of harmonic spectra
calculated for $O_2$, $N_2$ and $H_2^{+}$ depending on the problem
parameters proved to be in a reasonable or fairly good accordance
with relevant data calculated by earlier authors and observed in
molecular HHG experiments.}\\

\item ACKNOWLEDGMENTS.

\qquad {\rm This work was supported by the Chemical Sciences,
Geosciences and Biosciences Division of the Office of Basic Energy
Sciences, Office of Science, U.S. Department of Energy and also,
separately, by the Center of Sciences and Technology of Uzbekistan
(Project No. }$\Phi ${\rm -2.1.44). The support from DAAD
(Deutscher Akademischer Austauschdienst) and B-Division of
Max-Born-Institute of Nonlinear Optics and Short-Pulse Laser
Spectroscopy (Berlin, Germany) is also gratefully acknowledged.}

\qquad {\rm The research described in this presentation was made
also possible in part by financial support from the U.S. Civilian
Research and Development Foundation (CRDF) for the Independent
States of the Former Soviet Union.}
\end{enumerate}

\begin{center}
{\rm {\bf FIGURE\ CAPTIONS}}\\
\end{center}

\begin{enumerate}
\item  Fig.1. (Color online). The squared module $\left|
F_{3\sigma _g}\left({\bf p}_N{\bf ,R}_0\right) \right| ^2$ of
molecular wavefunction in momentum space (Fourier transform) for
$N_2$ vs the angle $\theta _{{\bf p}}$ (relative to the
internuclear molecular axis lined up in the horizontal direction)
of photoelectron emission from the $3\sigma _g$ HOMO of different
composition: (a) $\left( 2p\right) 3\sigma _g$ MO composed from
$2p_z$ states alone, (b) $ \left( 1s\right) 3\sigma _g$ MO
composed from $1s$ states alone, (c) $\left( 1s2p\right) 3\sigma
_g$ MO composed from $1s$ and $2p_z$ states, (d) $\left(
1s2s2p\right) 3\sigma _g$ MO composed from $1s$, $2s$ and $2p_z$
states. Those angular dependencies are presented for
photoelectrons only emitted along the laser field polarization due
to absorption of different number $N$ of laser photons beginning
from minimum one $N_0=18$ required for ionization of the $3\sigma
_g$ HOMO by {\it Ti:sapphire} incident laser field of the
wavelength $\lambda =800$ $nm$ ($\hbar \omega =1.55\ eV$) and
fixed intensity $I=2\cdot 10^{14}W/cm^2$ .

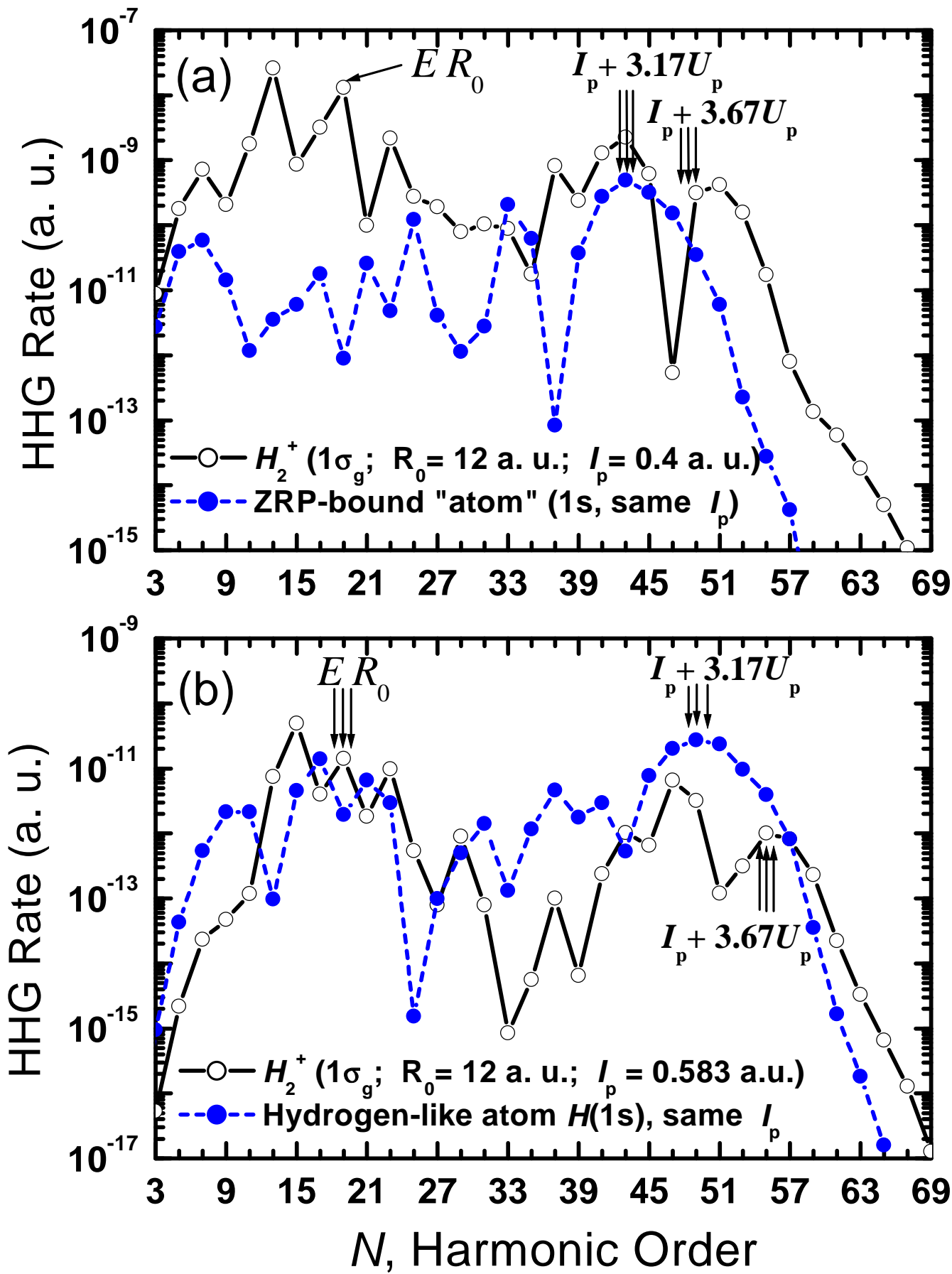
\item  Fig.2. (Color online). The differential harmonic emission
rates $w_N^{\left ( HHG\right) }({\bf k}^{\prime })$ [Eq.(13)] in
strongly aligned $H_2^{+}$ vs the harmonic order $N$ calculated
under conditions of Ref.[10] for incident laser field of the
wavelength $\lambda \approx 1064$ $nm$ ($\hbar \omega =1.172\ eV$)
and intensity $I=1.19\cdot 10^{14}$ $W/cm^2$ (so that $\eta
\approx 10.58$). The presented molecular high-harmonic spectra
(filled circles) are for the internuclear separation $R_0=12$
$a.u.$ assuming (a) the fixed (independent on $R_0$) binding
energy $\varepsilon_0$ ($\left| \varepsilon _0\right| =0.4$
$a.u.$) (b) $R_0$-dependent molecular binding energy $\varepsilon
_0\left( R_0\right)$ corresponding to $\left( 1s\right) 1\sigma_g$
valence shell. The respective spectra of model counterpart
''atom'' (open circles) of the same ionization potential $I_p$
were calculated assuming (a) the model zero-range binding
potential (ZRP), (b) pure Coulomb binding potential. The results
of Fig.2a are to be compared with relevant ones calculated within
a different HHG model [10] (see Fig.4 presented therein).

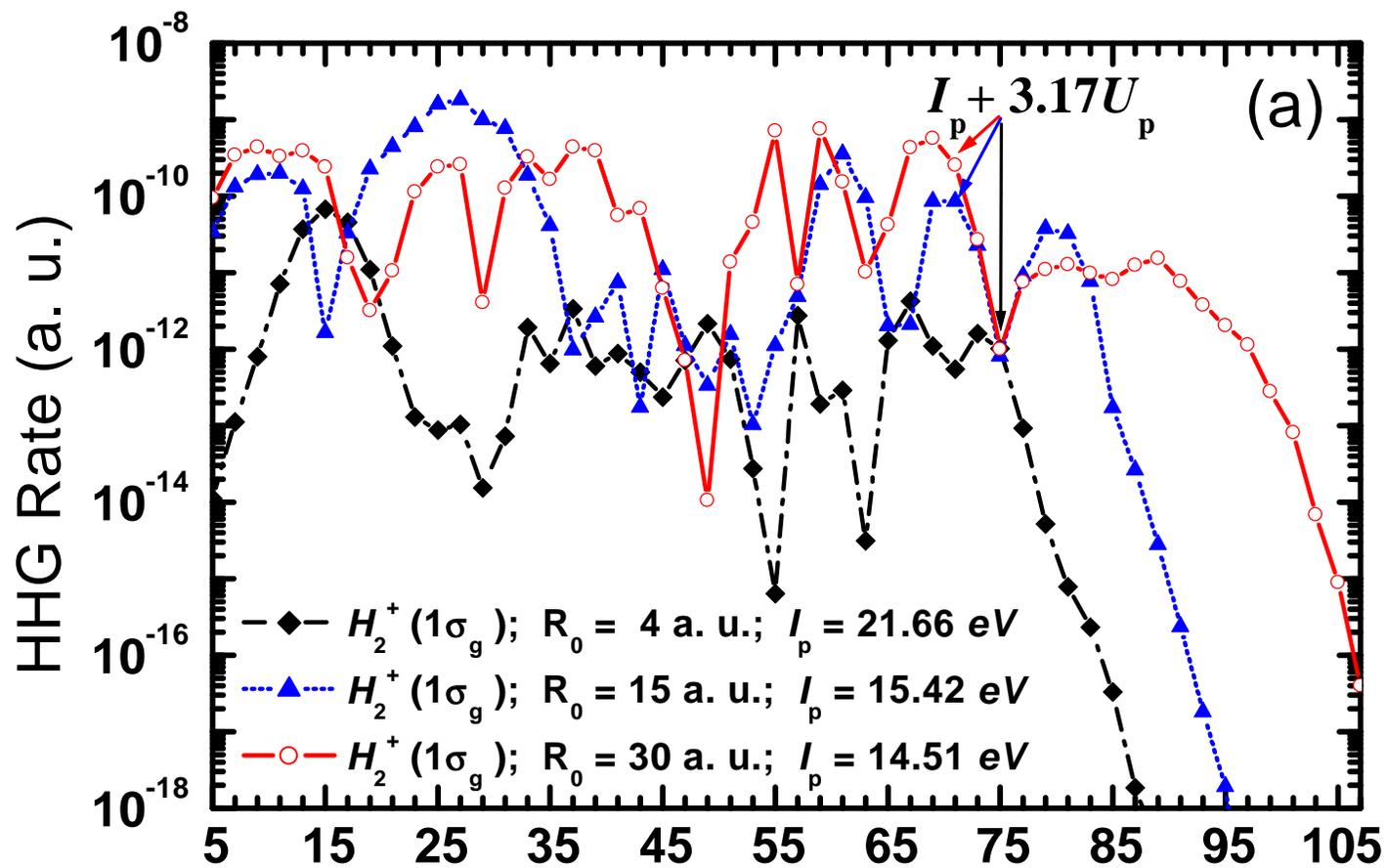
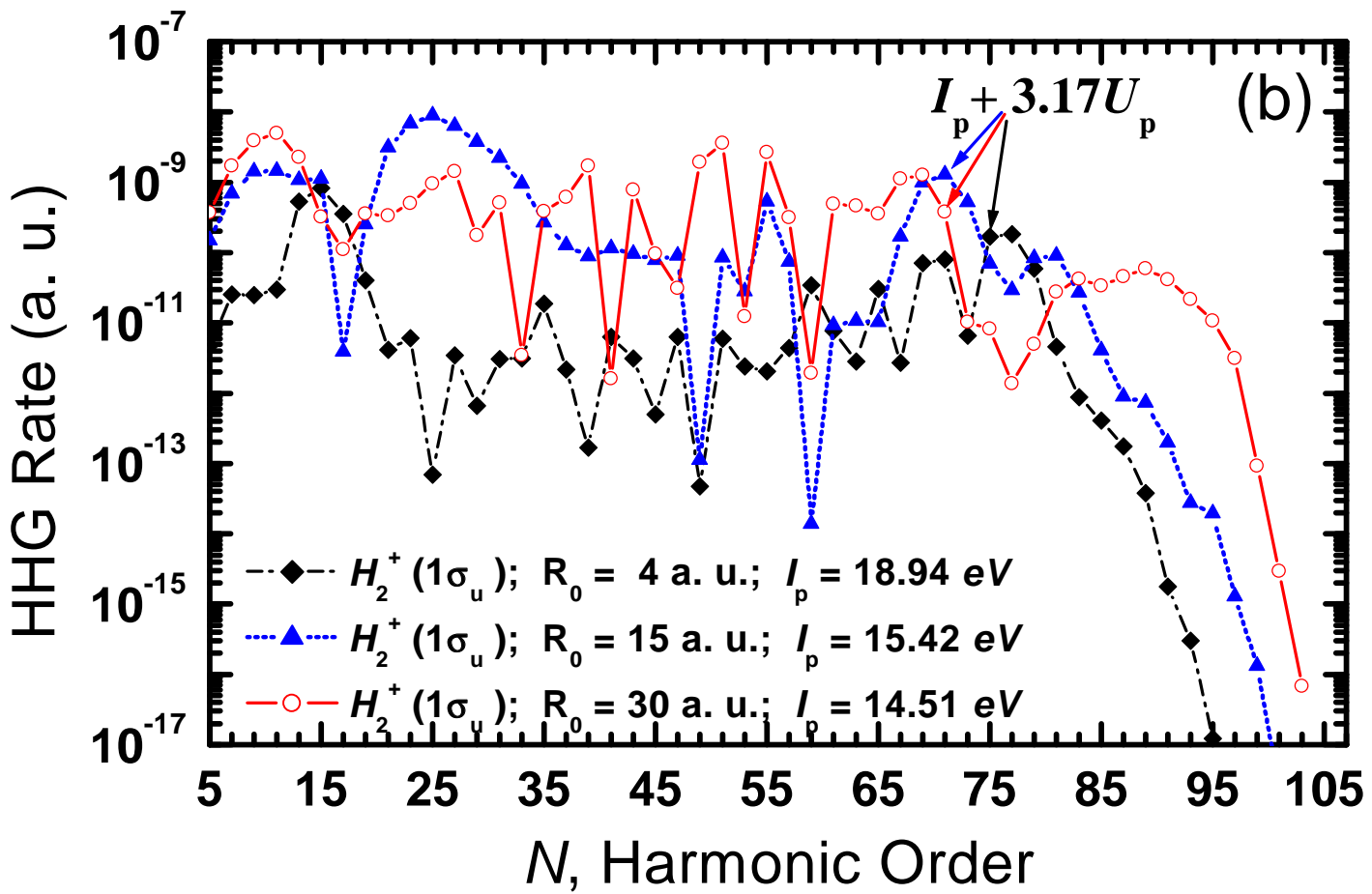
\item  Fig.3. (Color online). The differential harmonic emission
rates $w_N^{\left(HHG\right) }({\bf k}^{\prime })$ [Eq.(13)] vs
the harmonic order $N$ calculated for various internuclear
separations $R_0$ in $H_2^{+}$ strongly aligned along the
polarization of {\it Ti:sapphire} laser field of the wavelength
$\lambda \approx 800\ nm$ ($\hbar \omega =1.55\ eV$) and intensity
$I=5 \cdot 10^{14}W/cm^2$. The calculated harmonic spectra are
presented for two different bonding symmetry of initial molecular
state: (a) bonding $1\sigma _g$ molecular state, (b) antibonding
$1\sigma _u$ molecular state. The molecular binding energy
$\varepsilon _0$ is supposed to be dependent on internuclear
separation $R_0$, so that the dependence $ \varepsilon _0\left(
R_0\right) $ is to be different for $1\sigma _u$ and $ 1\sigma _g$
molecular orbitals. Beginning from $R_0=10$ $a.u.$ the dependence
$\varepsilon _0\left( R_0\right) $ becomes almost the same for $
1\sigma _u$ and $1\sigma _g$ valence shells and, to a fairly good
accuracy, can be approximately reproduced by the equation
$\varepsilon _0\left( R_0\right) \approx -0.5-1/R_0$. These
results are to be compared with relevant ones presented in
Ref.[11] (see, e.g., Fig.4 therein).

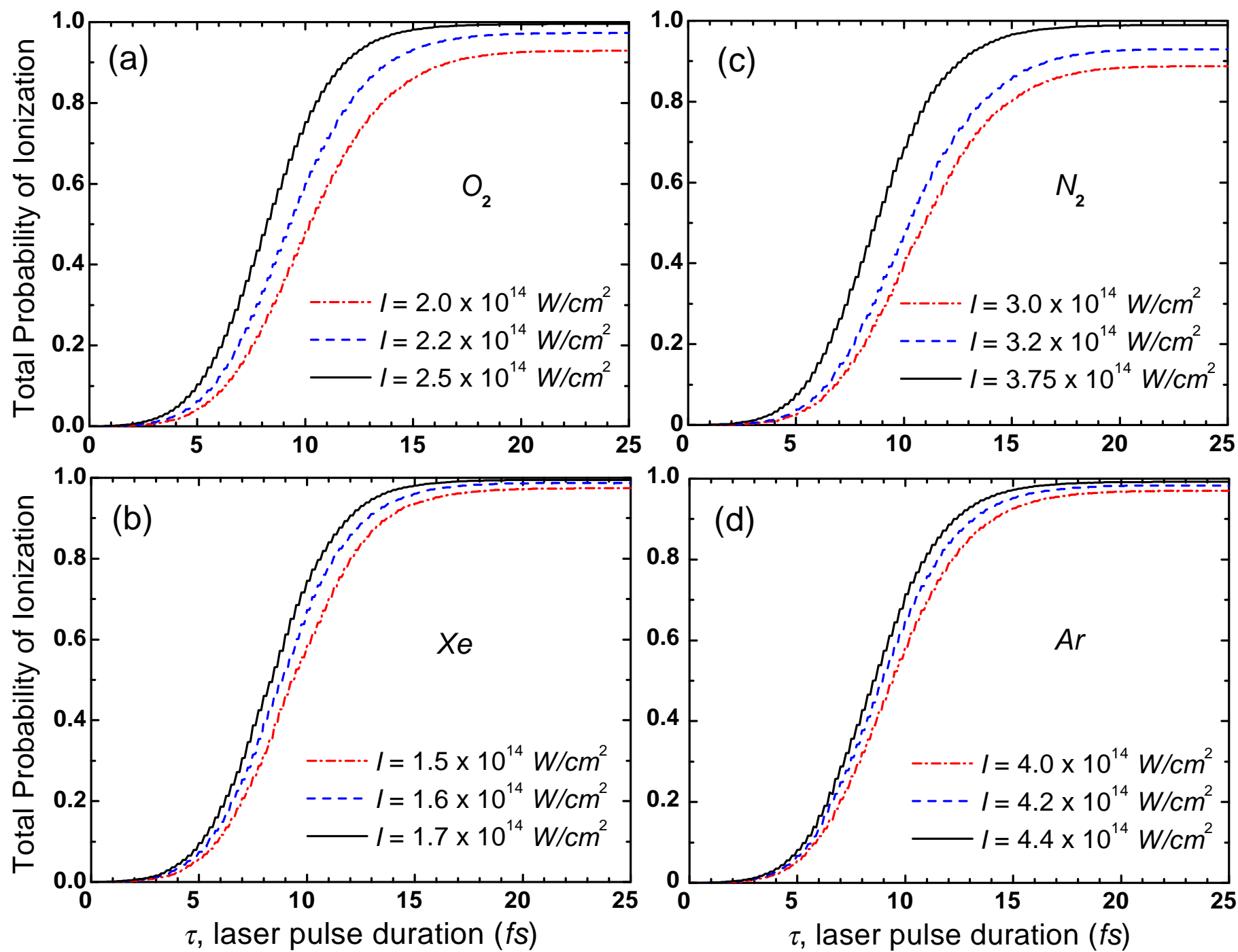
\item  Fig.4. (Color online). The temporal dependence of total
ionization probability $P_{ion}(\tau )$ [Eq.(17)] calculated under
conditions of experiment [13] for (a) diatomic $O_2$, (b) atomic
$Xe$, (c) diatomic $N_2$, (d) atomic $Ar$ irradiated by {\it
Ti:sapphire} laser field of the wavelength $\lambda \approx 800$
$nm$ and {\rm Gaussian temporal profile (16) with the }pulse
duration $\tau \approx 25$ $fs$ {\rm corresponding to about $9.43$
optical cycles. The presented results correspond to }various
values of laser peak intensity $I$: (a) $2.0\cdot 10^{14}$
$W/cm^2$ (dash-dotted line), $2.2\cdot 10^{14}$ $W/cm^2$ (dashed
line) and the saturation intensity $2.5\cdot 10^{14}$ $W/cm^2$
(solid line), (b) $1.5\cdot 10^{14}$ $W/cm^2$ (dash-dotted line),
$1.6\cdot 10^{14}$ $W/cm^2$ (dashed line) and the saturation
intensity $1.7\cdot 10^{14}$ $W/cm^2$ (solid line), (c) $3.4\cdot
10^{14}$ $ W/cm^2$ (dash-dotted line), $3.6\cdot 10^{14}$ $W/cm^2$
(dashed line) and the saturation intensity $3.75\cdot 10^{14}$
$W/cm^2$ (solid line), (d) $ 4.0\cdot 10^{14}$ $W/cm^2$
(dash-dotted line), $4.2\cdot 10^{14}$ $W/cm^2$ (dashed line) and
the saturation intensity $4.4\cdot 10^{14}$ $W/cm^2$ (solid line).

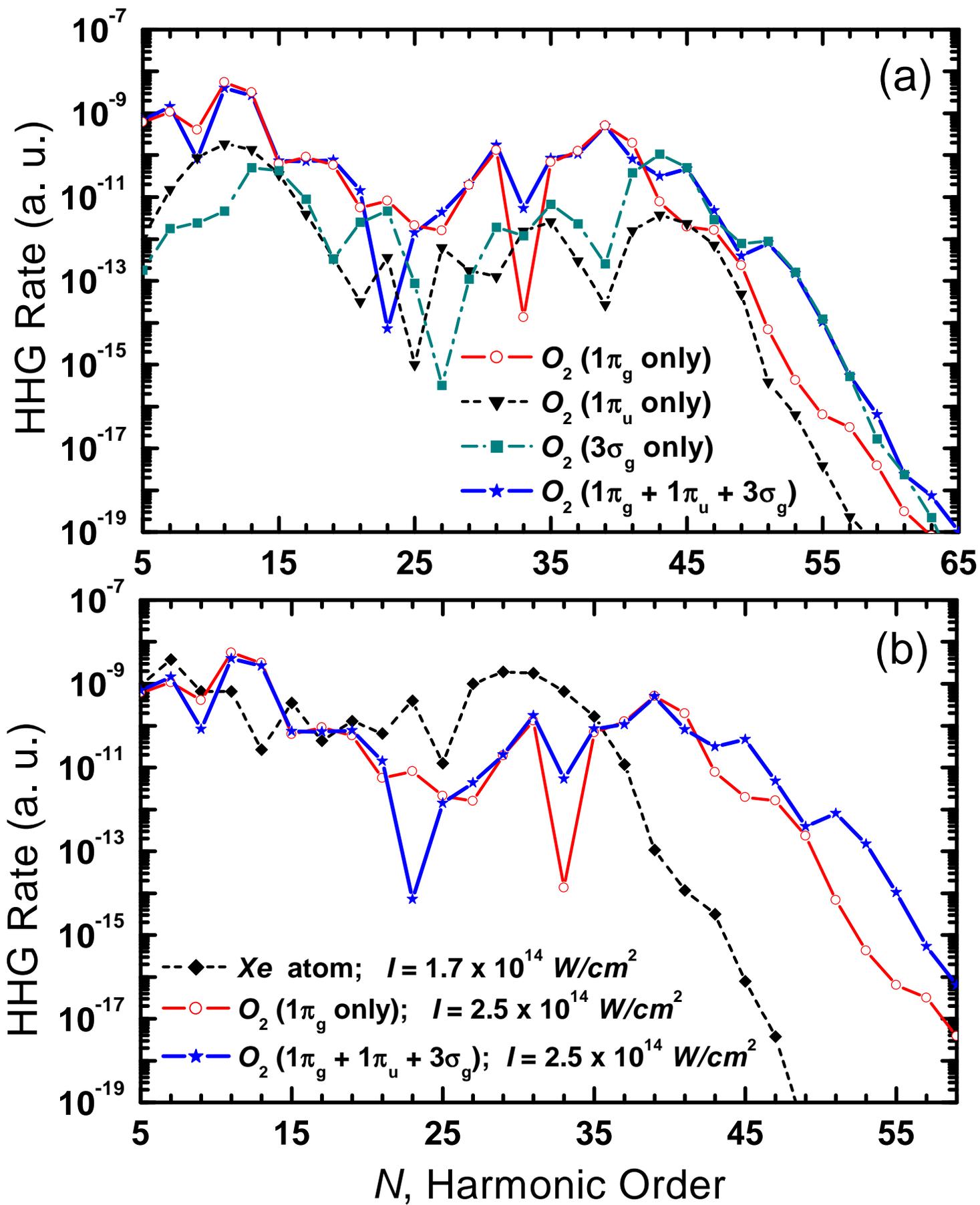
\item  Fig.5. (Color online). The differential harmonic emission
rates $w_N^{\left(HHG\right) }({\bf k}^{\prime })$ [Eq.(13)] in
laser-irradiated $O_2$ and its atomic counterpart $Xe$ vs the
harmonic order $N$ calculated under conditions of experiment [13]
for the laser pulse of respective saturation peak intensity
presently found and summarized in Table I. (a) The spectra of
$O_2$ corresponding to partial contributions from the $1\pi _g$
HOMO (open circles), the inner $1\pi _u$ (filled triangles) and
$3\sigma _g$ (filled squares) valence shells separately are
presented vs overall $O_2$ spectrum corresponding to coherent
contribution (32) from those three highest-lying molecular valence
shells under consideration put together (stars), (b) The overall
harmonic spectrum of $O_2$ presented vs the $Xe$ spectrum (filled
diamonds) calculated for respective (different from $O_2$)
saturation laser intensity. The results of Fig.5b are to be
compared with relevant ones presented in Fig.5 of Ref.[13].

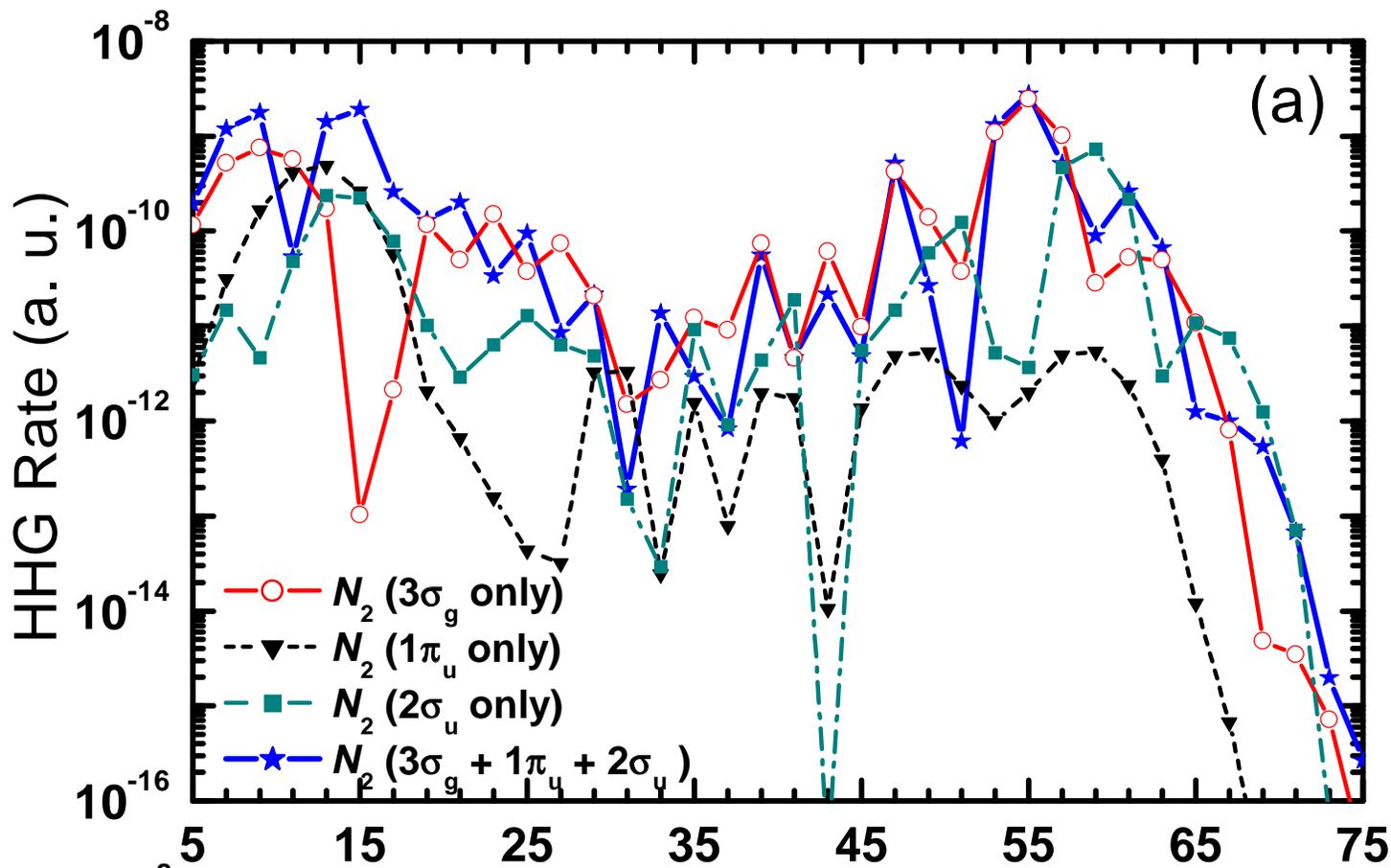
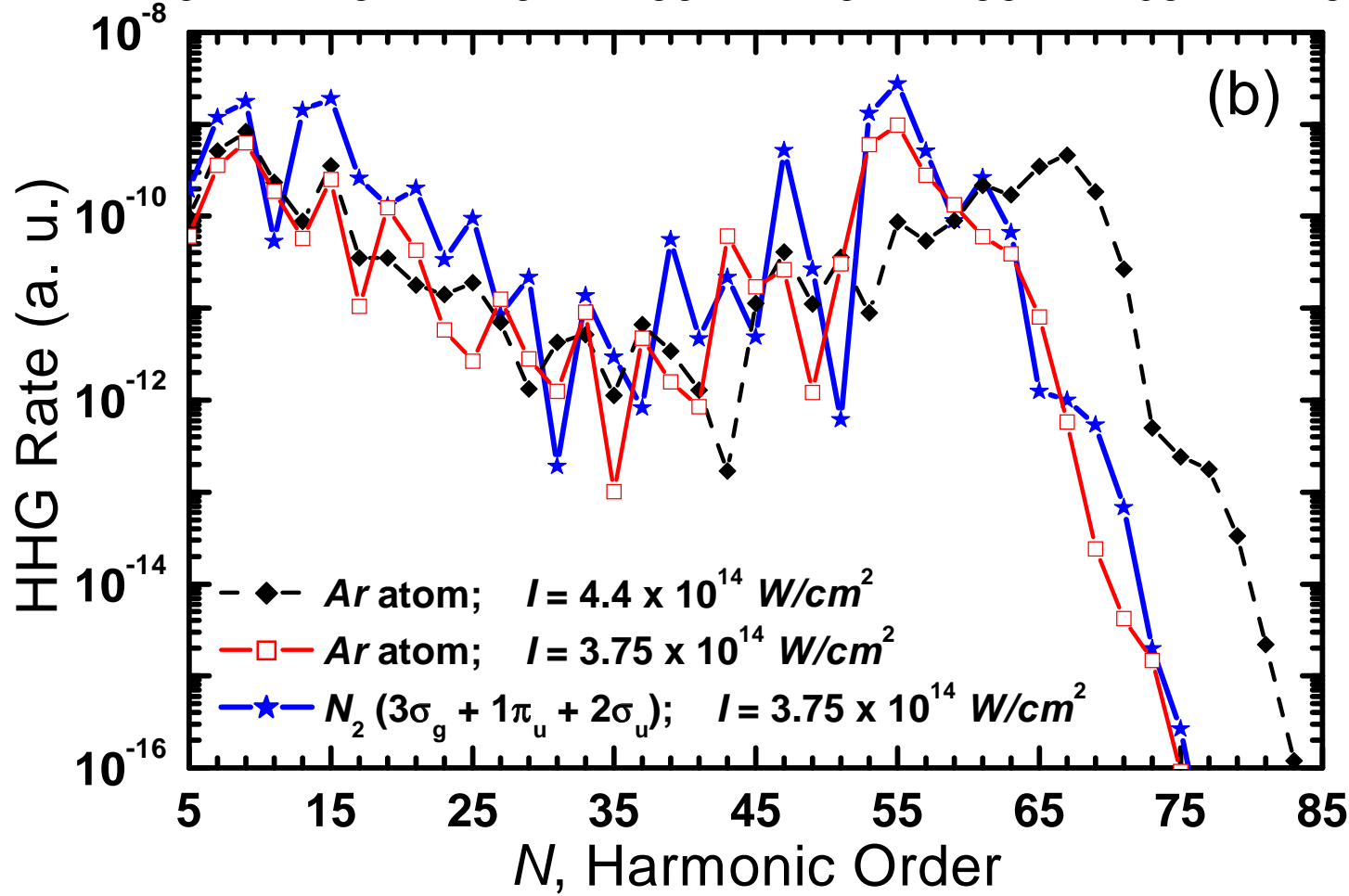
\item  Fig.6. (Color online). The differential harmonic emission
rates $w_N^{\left(HHG\right) }({\bf k}^{\prime })$ [Eq.(13)] in
strongly aligned $N_2$ and its atomic counterpart $Ar$ vs the
harmonic order $N$ calculated under conditions of experiment [13]
for the laser pulse of respective saturation peak intensity
presently found and summarized in Table I. (a) The spectra of
$N_2$ corresponding to partial contributions from the $3\sigma _g$
HOMO (open circles), the inner $1\pi _u$ (filled triangles) and
$2\sigma _u$ (filled squares) valence shells separately and
presented vs overall $N_2$ spectrum corresponding to coherent
contribution (31) from those three highest-lying molecular valence
shells under consideration put together (stars). (b) The overall
harmonic spectrum of $N_2$ (stars) presented vs two harmonic
spectra of $Ar$ calculated for two different laser peak
intensities - $4.4\cdot 10^{14}$ $W/cm^2$ (filled diamonds) and
$3.75\cdot 10^{14}$ $W/cm^2$ (open squares) corresponding to the
saturation intensities found for ionization of $Ar$ and $N_2$,
respectively The results of Fig.6b are to be compared with
relevant ones presented in Fig.5 of Ref.[13].
\end{enumerate}

\end{document}